\documentclass[pr,onecolumn,10pt]{revtex4-2}
\usepackage{amsmath,amssymb,graphicx,subfigure}
\usepackage{times}
\usepackage{subfigure}

\begin{document}

\title{One-dimensional Townes solitons in dual-core systems with localized
coupling }
\author{ Shatrughna Kumar$^{1}$, Pengfei Li$^{2,3}$, and Boris A. Malomed$%
^{1,4}$}
\address{$^{1}$Department of Physical Electronics, School of Electrical
Engineering, Faculty of Engineering, and Center for Light-Matter
Interaction, Tel Aviv University, P.O.B. 39040, Tel Aviv, Israel\\
$^{2}$Department of Physics, Taiyuan Normal University, Jinzhong, 030619,
China\\
$^{3}$Institute of Computational and Applied Physics, Taiyuan Normal
University, Jinzhong, 030619, China\\
$^4$Instituto de Alta Investigaci\'{o}n, Universidad de Tarapac\'{a},
Casilla 7D, Arica, Chile}

\begin{abstract}
The recent creation of Townes solitons (TSs) in binary Bose-Einstein
condensates \cite{TS-exper,TS-exper 2} and experimental demonstration of
spontaneous symmetry breaking (SSB) in solitons propagating in dual-core
optical fibers \cite{Bugar} draw renewed interest to the TS and SSB
phenomenology in these and other settings. In particular, stabilization of
TSs, which are always unstable in free space, is a relevant problem with
various ramifications. We introduce a system which admits exact solutions
for both TSs and SSB of solitons. It is based on a dual-core waveguide with
quintic self-focusing and fused (localized) coupling between the cores. The
respective system of coupled nonlinear Schr\"{o}dinger equations gives rise
to exact solutions for full families of symmetric solitons and asymmetric
ones, which are produced by the supercritical SSB bifurcation (i.e., the
symmetry-breaking phase transition of the second kind). Stability boundaries
of asymmetric solitons are identified by dint of numerical methods. Unstable
solitons spontaneously transform into robust moderately asymmetric breathers
or strongly asymmetric states with small intrinsic oscillations. The setup
can be used in the design of photonic devices operating in coupling and
switching regimes.
\end{abstract}

\maketitle

\section{Introduction}

Critical collapse is a fundamental phenomenon with well-known manifestations
in optics, plasmas, Bose-Einstein condensates (BECs), etc. \cite%
{Berge,Sulem,Zakh-Kuz,Fibich,book}. Inherently related to it is the notion
of Townes solitons (TSs), which were first predicted (before the concept of
solitons was introduced by Zabuski and Kruskal, and the term was coined by
them \cite{solitons}) as two-dimensional (2D) light beams propagating in an
optical medium with the cubic self-focusing \cite{Townes}. TSs play the role
of boundaries (separatrices) between collapsing and decaying dynamical
states.

These concepts are adequately represented by the ubiquitous nonlinear Schr\"{%
o}dinger (NLS) equations for the wave amplitude $\psi (\mathbf{r},z)$,
\begin{equation}
i\psi _{z}=-\frac{1}{2}\nabla ^{2}\psi -|\psi |^{n-1}\psi ,  \label{NLS}
\end{equation}%
where, in terms of optics, $z$ and $\mathbf{r}$ are the propagation distance
and set of transverse coordinates in the $D$-dimensional space, and $n$ is
the power of the self-focusing nonlinearity. This equation can be derived
from the Hamiltonian,%
\begin{equation}
H=\int \left( \frac{1}{2}\left\vert \nabla \psi \right\vert ^{2}-\frac{2}{n+1%
}|\psi |^{n+1}\right) d^{D}\mathbf{r.}  \label{H}
\end{equation}%
To analyze the possibility of the onset of the collapse governed by Eq. (\ref%
{NLS}), one can consider a localized state of size $L$ with amplitude $A$.
The conservation of the total power (norm), $P=\int \left\vert \psi (\mathbf{%
r})\right\vert ^{2}d^{D}\mathbf{r}$, imposes a relation on the scaling of $A$
and $L$ in the course the development of the collapse, which corresponds to
the evolution towards $L\rightarrow 0$: $A\sim \sqrt{P}L^{-D/2}$. Taking
this relation into account, a straightforward estimate shows that the
gradient and self-attraction terms in the Hamiltonian scale, at $%
L\rightarrow 0$, as
\begin{equation}
H_{\mathrm{grad}}\sim +PL^{-2},H_{\mathrm{self-attr}}\sim
-P^{(n+1)/2}L^{-(D/2)(n-1)},  \label{HH}
\end{equation}%
The comparison between these terms leads to the conclusion that the negative
term dominates and leads to the \textit{supercritical collapse} at $D(n-1)>4$%
, starting from the input with arbitrarily small $P$. In the opposite case, $%
D(n-1)<4$, the positive term is the dominant one in Eq. (\ref{HH}), thus
preventing the collapse and making it possible to construct stable solitons.
At the border, i.e., at
\begin{equation}
D(n-1)=4,  \label{D}
\end{equation}%
the \textit{critical collapse}, leading to emergence of a singularity in the
solution, takes place after a finite propagation distance. In this case, the
negative term dominates in Eq. (\ref{HH}), which happens if $P$ exceeds a
critical value, $P_{\mathrm{cr}}$, while at $P<P_{\mathrm{cr}}$ the positive
term is dominating in Eq. (\ref{HH}), leading to decay of the input. In the
case of $D=2$ and $n=3$, the occurrence of the singularity at $P>P_{\mathrm{%
cr}}$ is a rigorous corollary of the virial theorem \cite%
{virial,Berge,Fibich}.

Stationary solutions to Eq. (\ref{NLS}), $\psi (\mathbf{r},z)=\exp \left(
ikz\right) u(\mathbf{r})$, with propagation constant $k$, are invariant with
respect to the conformal transformation,%
\begin{equation}
\mathbf{r}\rightarrow \rho \mathbf{r},\psi \rightarrow \rho ^{-2/(n-1)}\psi
,k\rightarrow \rho ^{-2}k,P\rightarrow \rho ^{D-4/(n-1)}P,  \label{scaling}
\end{equation}%
where $\rho $ is an arbitrary scaling factor. In the critical case (\ref{D}%
), Eq. (\ref{scaling}) demonstrates that the soliton family, which is one of
the TS type, is degenerate, as its power takes a single value that does not
depend on $k$ (e.g., $P_{\mathrm{TS}}\approx 5.85$ for $D=2$ and $n=3$, its
counterpart predicted by the variational approximation being $P_{\mathrm{TS}%
}=2\pi $ \cite{Anderson}).

As mentioned above, TSs play the role of separatrices between the collapsing
and decaying solutions in the case when critical condition (\ref{D}) holds.
For this reason, TSs are unstable against small perturbations, although the
initial growth of small perturbations is subexponential, accounted for by a
pair of zero eigenvalues in the respective spectrum \cite%
{Berge,Sulem,Zakh-Kuz,Fibich}. The latter circumstance suggests that the
instability grows slowly at the initial stage, making it recently possible
to experimentally observe 2D\ TSs in a binary BEC \cite{TS-exper,TS-exper 2}%
, which is modeled by the Gross-Pitaevskii equation in the form of Eq. (\ref%
{NLS}) with $n=3$. This\emph{\ first experimental demonstration} of physical
states of the TS type, achieved 57 years after they had been predicted \cite%
{Townes}, makes further studies of TSs in other physical settings a relevant
objective. In particular, a remaining challenge is to elaborate
possibilities for observing such modes in optics, where they were predicted
for the first time.

In the 1D case, the critical collapse and TS family correspond to $n=5$ in
Eq. (\ref{D}), i.e., the quintic self-focusing \cite{1D-TS,1D-TS-2}:

\begin{equation}
i\psi _{z}=-\frac{1}{2}\psi _{xx}-|\psi |^{4}\psi ,  \label{quintic}
\end{equation}%
where $x$ is the transverse coordinate in the planar waveguide. The quintic
nonlinearity occurs in various optical materials, often in a combination
with a cubic term. The effective coefficients before quintic and cubic terms
can be accurately adjusted to target values, including the case of the
quintic-only nonlinearity (nullifying the cubic coefficient), for the light
propagation through colloidal suspensions of metallic nanoparticles \cite%
{Cid-OptExp,Cid}. This possibility is provided by selecting an appropriate
density of the nanoparticles and their size, as well as the wavelength of
light. In terms of the light propagation in optical fibers, an equation
similar to Eq. (\ref{quintic}), with $x$ replaced by the reduced temporal
variable $\tau \equiv t-z/V_{\mathrm{gr}}$ (here $t$ is time, and $V_{%
\mathrm{gr}}$ is the group velocity of the carrier wave), may also give rise
to the critical collapse \cite{Berge12}. However, the collapse in optical
fibers is easily suppressed by higher-order effects, such as the stimulated
Raman scattering \cite{Hasegawa}.
%In the realm of quantum matter, Eq. (\ref%
%{quintic}), with $z$ replaced by time, may serve as an approximation for
%super-Tonks-Girardeau gases \cite{TG1,TG2}.

The family of TS solutions to Eq. (\ref{quintic}) is commonly known,%
\begin{equation}
\psi _{\mathrm{TS}}=\exp \left( ikz\right) \frac{\left( 3k\right) ^{1/4}}{%
\sqrt{\cosh \left( \sqrt{8k}x\right) }},  \label{simple}
\end{equation}%
where real propagation constant $k$ takes values in the semi-infinite
interval, $0<k<\infty $. In accordance with what is said above, the family
is degenerate, as the power of all the solutions takes the single value,
which does not depend of $k$:
\begin{equation}
P_{\mathrm{TS}}=\int_{-\infty }^{+\infty }\left\vert \psi \left( x\right)
\right\vert ^{2}dx=\frac{\sqrt{6}}{4}\pi \approx 1.92.  \label{NTS}
\end{equation}%
The well-known necessary stability condition for soliton families, if they
are characterized by the dependence of the power on the propagation
constant, is the \textit{Vakhitov-Kolokolov (VK) criterion} \cite%
{VK,Berge,Sulem,Zakh-Kuz,Fibich}, $dP/dk>0$. Generally, it is only necessary
but not sufficient for the stability of the solitons. For the degenerate TS
family, with $dP/dk=0$, the VK\ criterion yields a neutral result. Actually,
as mentioned above, TSs are always unstable, while the formally neutral VK
prediction corresponds to the above-mentioned fact that their instability
against small perturbations is subexponential, corresponding to the pair of
zero eigenvalues for small perturbations.

Dual-core waveguides, alias \textit{couplers}, are systems which realize
another fundamental effect, \textit{viz}., spontaneous symmetry breaking
(SSB) -- in particular, in optics \cite{APL,Wright,Snyder,Wabnitz,review}
and BEC\ \cite{Smerzi,Markus}. In the latter setting, SSB was experimentally
realized about 20 years ago \cite{Markus}, while SSB in nonlinear optical
couplers, predicted still earlier \cite{Wright}, was demonstrated
experimentally for solitons in dual-core fibers 31 years later \cite{Bugar}.

The well-elaborated theoretical predictions for the critical collapse and
SSB, and the recent experimental works which have demonstrated these
phenomena separately \cite{TS-exper,Bugar}, suggest a possibility to look
for new physical effects produced by their interplay in dual-core
waveguides. In particular, it is interesting to use such systems as a means
for the stabilization of TSs. It is demonstrated below, by means of an \emph{%
exact solution}, that this is possible indeed. The exact solution offers a
reference point for the study of stabilized TS modes in a more general form.

In the simplest case, the system combining the critical collapse and SSB is
based on the system of linearly coupled NLS equations with the quintic
intra-core nonlinearity,

\begin{eqnarray}
i\psi _{z} &=&-\frac{1}{2}\psi _{xx}-|\psi |^{4}\psi -\lambda \phi ,
\label{1} \\
i\phi _{z} &=&-\frac{1}{2}\phi _{xx}-|\phi |^{4}\phi -\lambda \psi ,
\label{2}
\end{eqnarray}%
where $\lambda >0$ is a real coupling constant. It can be realized
experimentally as a set of two parallel planar optical waveguides with the
nonlinearity adjusted to the quintic form as outlined above.
%Alternatively,
%Eqs. (\ref{1}) and (\ref{2}), with $z$ replaced by time, may serve as an
%approximate model for the super-Tonks-Gigardeau gas loaded in a
%tunnel-coupled pair of parallel quasi-1D traps.

Soliton solutions to Eqs. (\ref{1}) and (\ref{2}) with propagation constant $%
k$ are looked for as
\begin{equation}
\left\{ \psi (x,z),\phi (x,z)\right\} =\exp \left( ikz\right) \left\{
u(x),v(x)\right\} ,  \label{u,v}
\end{equation}%
where spatially even real functions $u(x)$ and $v(x)$ satisfy stationary
equations%
\begin{eqnarray}
-ku &=&-\frac{1}{2}\frac{d^{2}u}{dx^{2}}-u^{5}-\lambda v,  \label{u} \\
-kv &=&-\frac{1}{2}\frac{d^{2}v}{dx^{2}}-v^{5}-\lambda u,  \label{v}
\end{eqnarray}%
with boundary conditions $u(|x|\rightarrow \infty )=v(|x|\rightarrow \infty
)=0$. The solitons are characterized by powers of their two components,%
\begin{equation}
P_{u}\equiv \int_{-\infty }^{+\infty }u^{2}(x)dx,P_{v}\equiv \int_{-\infty
}^{+\infty }v^{2}(x)dx,  \label{NN}
\end{equation}%
and the total power, $P\equiv P_{u}+P_{v}$, the latter one being a dynamical
invariant of the system.

It is well known that, in the case of the cubic self-focusing, with terms $%
|\psi |^{4}\psi $ and $|\phi |^{4}\phi $ in Eqs. (\ref{1}) and (\ref{2})
replaced by $|\psi |^{2}\psi $ and $|\phi |^{2}\phi $, respectively, the
competition between the intra-core self-focusing and linear inter-core
coupling leads to the SSB bifurcation (alias symmetry-breaking phase
transition), which destabilizes obvious symmetric soliton solutions, with $%
\psi =\phi $, and replaces them by stable asymmetric ones \cite%
{Wabnitz,review} when $P$ and $k$ exceed the corresponding critical values,
\begin{equation}
P_{\mathrm{SSB}}^{\mathrm{(cubic)}}=8\sqrt{\lambda }/3,k_{\mathrm{SSB}}^{%
\mathrm{(cubic)}}=5\lambda /3.  \label{PK}
\end{equation}%
For the system of Eqs. (\ref{1}) and (\ref{2}), a similar analysis
demonstrates that the SSB takes place at $k\geq k_{\mathrm{SSB}}^{\mathrm{%
(quintic)}}=5\lambda /4$. On the other hand, all the symmetric solitons
produced by Eqs. (\ref{1}) and (\ref{2}) assume a single value of the power,
which is $2P_{\mathrm{TS}}=\pi \sqrt{3/2}$, according to Eq. (\ref{NTS}),
therefore there is no special value of $P_{\mathrm{SSB}}^{\mathrm{(quintic)}%
} $. Finally, the analysis of the Hamiltonian of the system of Eqs. (\ref{1}%
) and (\ref{2}), similar to that leading to Eq. (\ref{HH}), leads to the
conclusion that the system with the quintic self-focusing and simplest
linear coupling can only create unstable solitons of the TS type.

The objective of this work is to introduce a new system of linearly coupled
NLS equations with the quintic self-focusing, which admits exact stable
solutions for symmetric and asymmetric solitons (thus, the system provides
an exact solution for the SSB phase transition). Unlike the system of Eqs. (%
\ref{1}) and (\ref{2}), the new one includes localized coupling (implying
that a \textit{fused} \cite{fused2,fused1} dual-core system is considered),
and also a local attractive potential attached to the coupling spot, which
helps to stabilize the solitons in the system. The model is introduced, with
necessary details, in Section II. The exact solutions for solitons, which
are symmetric and asymmetric with respect to the coupled cores, are reported
in Section III. The analysis of the solitons' stability makes it necessary
to use numerical methods, with the results summarized in Section IV. In
particular, it is found that asymmetric solitons, created by the SSB, are
stable in a finite interval of values of the propagation constant. Evolution
of those symmetric and asymmetric states which are unstable leads to
establishment of either moderately asymmetric breathers (periodically
oscillating localized modes) or strongly asymmetric states with small
intrinsic vibrations. The work is concluded by Section V, where, in
particular, we provide estimates of physical parameters for the experimental
realization of the system, and solitons maintained by it, in optical
couplers.

\section{The model}

\subsection{Local attractive potentials}

\subsubsection{The linear potential}

A possibility for the stabilization of TSs in the 1D single-component NLS
equation with the quintic self-focusing was elaborated in Ref. \cite{LiWang}%
, where a delta-functional attractive potential (local defect) was added to
Eq. (\ref{quintic}):%
\begin{equation}
i\psi _{z}=-\frac{1}{2}\psi _{xx}-|\psi |^{4}\psi -\varepsilon \delta
(x)\psi ,  \label{defect}
\end{equation}%
with potential strength $\varepsilon >0$. In optics, it may be realized as a
narrow stripe with a higher refractive index embedded in the planar
waveguide. A family of exact solutions for solitons pinned to the attractive
defect is%
\begin{equation}
\psi \left( x,z\right) =\exp \left( ikz\right) \frac{\left( 3k\right) ^{1/4}%
}{\sqrt{\cosh \left( 2\sqrt{2k}(|x|+\xi \right) }},  \label{delta}
\end{equation}%
cf. Eq. (\ref{simple}), with offset parameter $\xi >0$ determined by equation%
\begin{equation}
\tanh \left( 2\sqrt{2k}\xi \right) =\varepsilon /\sqrt{2k}.  \label{tanh}
\end{equation}%
Because $\tanh $ takes only values $<1$, the solution for the pinned
solitons, admitted by Eq. (\ref{tanh}), exists in interval%
\begin{equation}
\varepsilon ^{2}/2<k<\infty .  \label{1/2}
\end{equation}

For the soliton family given by Eqs. (\ref{delta}) and (\ref{tanh}), the
dependence of the power on the propagation constant is \cite{LiWang}%
\begin{equation}
P(k)=\sqrt{\frac{3}{2}}\left[ \pi -2\arctan \left( \sqrt{\frac{\sqrt{2k}%
+\varepsilon }{\sqrt{2k}-\varepsilon }}\right) \right] .  \label{N(mu)}
\end{equation}%
This expression starts from $P=0$ at the left edge of the existence range (%
\ref{1/2}), $k=\varepsilon ^{2}/2$, and attains the limit value (\ref{NTS})
in the limit of $k\rightarrow \infty $. Thus, the attractive
delta-functional potential lifts the power degeneracy of the TS family (\ref%
{simple}).

Dependence $P(k)$ given by Eq. (\ref{N(mu)}) satisfies the VK criterion. In
agreement with this fact, it was found, by means of numerical methods, that
the pinned-soliton family is completely stable \cite{LiWang}. The same
result can be predicted by the estimate of the Hamiltonian, following the
pattern of Eq. (\ref{HH}): the estimate of the extra term which accounts for
the attractive potential in Eq. (\ref{defect}), $H_{\varepsilon
}=-\varepsilon \left\vert \psi (x=0)\right\vert ^{2}$, yields $%
H_{\varepsilon }\sim -P/L$, cf. Eq. (\ref{HH}). Then, at $P<P_{\mathrm{TS}}$
[note that Eq. (\ref{N(mu)}) indeed yields values of $P$ which are smaller
than $P_{\mathrm{TS}}$ given by Eq. (\ref{NTS})], the analysis of the
estimate for $H$, including term $H_{\varepsilon }$, predicts a minimum of
the Hamiltonian at a finite value of $L$, which corresponds to stable
solitons.

\subsubsection{The nonlinear potential}

For the sake of comparison with the model based on Eq. (\ref{defect}), it is
relevant to mention one in which the attractive defect is not linear, but
cubic, which was introduced in Ref. \cite{Azbel}:%
\begin{equation}
i\phi _{z}=-\frac{1}{2}\phi _{xx}-\delta (x)|\phi |^{2}\phi .
\label{delta-cubic}
\end{equation}%
In optics, it describes a planar waveguide with a narrow strip of strong
local nonlinearity, which may be built by the respective distribution of the
density of dopants that induce the local self-focusing \cite{Kip}. In terms
of BEC, Eq. (\ref{delta-cubic}), with $z$ replaced by $t$, is the
Gross-Pitaevskii equation with the local self-attractive nonlinearity, which
may be imposed by a tightly focused laser beam through the optically-induced
Feshbach resonance \cite{Feshbach1,Feshbach2}. Equation (\ref{delta-cubic})
gives rise to a family of pinned states,%
\begin{equation}
\phi =\left( 2k\right) ^{1/4}\exp \left( ikz-\sqrt{2k}|x|\right) ,
\label{pinned}
\end{equation}%
whose power does not depend on $k$, \textit{viz}., $P=1$, i.e., Eq. (\ref%
{delta-cubic}) is another 1D model which gives rise to the TS family [hence
solutions (\ref{pinned}) are unstable].

Further, it is possible to consider the 1D NLS equation which combines
nonlinear terms from both equations (\ref{quintic}) and (\ref{delta-cubic})
which give rise to the critical collapse, namely,
\begin{equation}
i\phi _{z}=-\frac{1}{2}\phi _{xx}\mp |\phi |^{4}\phi -\varepsilon \delta
(x)|\phi |^{2}\phi .  \label{cubic-delta}
\end{equation}%
Unlike Eq. (\ref{defect}), coefficient $\varepsilon >0$ in Eq. (\ref%
{cubic-delta}) cannot be eliminated by rescaling. Exact solutions to Eq. (%
\ref{cubic-delta}) for pinned solitons take the form [cf. Eq. (\ref{delta})]%
\begin{equation}
\left\{
\begin{array}{c}
\phi _{-}\left( x,z\right) \\
\phi _{+}\left( x,z\right)%
\end{array}%
\right\} =\left( 3k\right) ^{1/4}\exp \left( ikz\right) \left\{
\begin{array}{c}
1/\sqrt{\cosh \left( 2\sqrt{2k}(|x|+\xi _{\_}\right) } \\
1/\sqrt{\sinh \left( 2\sqrt{2k}(|x|+\xi _{+}\right) }%
\end{array}%
\right\} ,  \label{-+}
\end{equation}%
with offset $\xi _{\mp }$ defined by equation [cf. Eq. (\ref{tanh})]
\begin{equation}
\left\{ \sinh \left( 2\sqrt{2k}\xi _{-}\right) ,\cosh \left( 2\sqrt{2k}\xi
_{+}\right) \right\} =\sqrt{\frac{3}{2}}\varepsilon .  \label{sinh}
\end{equation}%
Here subscripts $-$ and $+$ correspond to the same signs in Eq. (\ref%
{cubic-delta}), where they represent the self-focusing/defocusing bulk
nonlinearity, respectively. Note that, in the latter case, Eq. (\ref{sinh})
produces a solution if the localized cubic self-focusing is stronger than
the bulk defocusing, \textit{viz}., $\varepsilon >\sqrt{2/3}$. The
calculation of the power for these soliton families yields%
\begin{gather}
P_{-}(\varepsilon )=\sqrt{6}\left[ \frac{\pi }{2}-\arctan \left( \sqrt{\frac{%
3}{2}}\varepsilon +\sqrt{\frac{3}{2}\varepsilon ^{2}+1}\right) \right] ;
\notag \\
P_{+}(\varepsilon )=\frac{1}{2}\sqrt{\frac{3}{2}}\ln \left( \frac{%
\varepsilon +\sqrt{2/3}}{\varepsilon -\sqrt{2/3}}\right) .  \label{Ncubic}
\end{gather}%
Both expressions (\ref{Ncubic}) again do not depend on $k$, i.e., the
soliton families produced by Eqs. (\ref{-+}) and (\ref{sinh}), with the
combination of two nonlinear terms, also belong to the class of TS
solutions, hence they are unstable too.

\subsection{The system with fused coupling}

As suggested by Refs. \cite{Raymond,Alon}, stable two-component solitons can
be supported by a system with localized (\textit{fused }\cite{fused2,fused1}%
) optical coupling, which is modeled by equations%
\begin{eqnarray}
i\psi _{z} &=&-\frac{1}{2}\psi _{xx}-|\psi |^{4}\psi -\delta (x)\phi
-\varepsilon \delta (x)\psi ,  \label{eps1} \\
i\phi _{z} &=&-\frac{1}{2}\phi _{xx}-|\phi |^{4}\phi -\delta (x)\psi
-\varepsilon \delta (x)\phi ,  \label{eps2}
\end{eqnarray}%
cf. the model with the uniform coupling, provided by Eqs. (\ref{1}) and (\ref%
{2}). The fused coupling, approximated by the delta-functional terms with
the coefficient scaled to be $1$ in Eqs. (\ref{eps1}) and \ref{eps2}), can
be realized as a bridge connecting two parallel waveguiding cores \cite%
{Raymond,Alon}. The bridge also introduces a local increase of the thickness
in each core, which may be modeled, in the simplest form, by the
delta-functional potential terms with coefficient $\varepsilon >$ $0$ (the
latter terms were not considered in Refs. \cite{Raymond,Alon}). The local
attractive potential draws the wave field to the spot at which the coupler
is installed, thus helping to strengthen the coupling effect. In the
fused-coupler setup, the attractive potential can be enhanced by a locally
infused resonant dopant.

In addition to its potential use in photonic devices operating in the
coupling regime \cite{Chen,couplers}, the system based on Eqs. (\ref{eps1})
and (\ref{eps2}) offers a benefit for the theoretical study, as it provides
exact analytical solutions for the symmetry-breaking phenomenology of TS
families, as shown in the next section. Further, the analytical solutions
provide clues for the understanding of the operation of a realistic setup,
in which the ideal delta-function is replaced by its finite-width
regularization, as is demonstrated below in the numerical part of the work.

%In terms of the super-Tonks-Girardeau gas, the localized coupling can be
%readily created by a red-detuned laser beam illuminating the parallel traps
%in the perpendicular direction. Attracting atoms from both traps, it
%naturally gives rise to both delta-functional terms in Eqs. (\ref{eps1}) and
%(\ref{eps2}).

\section{Analytical solutions}

\subsection{The solution ansatz}

Stationary solutions to Eqs. (\ref{eps1}) and (\ref{eps2}), with real
propagation constant $k$, are looked for as per Eq. (\ref{u,v}), with real
functions $u(x)$ and $v(x)$ satisfying equations%
\begin{eqnarray}
-ku &=&-\frac{1}{2}\frac{d^{2}u}{dx^{2}}-u^{5}-\delta (x)\left(
v+\varepsilon u\right) ,  \label{udelta} \\
-kv &=&-\frac{1}{2}\frac{d^{2}v}{dx^{2}}-v^{5}-\delta (x)\left(
u+\varepsilon v\right) ,  \label{vdelta}
\end{eqnarray}%
cf. Eqs. (\ref{u}) and (\ref{v}). The solution to Eq. (\ref{defect})
represented by expression (\ref{delta}) suggests an ansatz for exact
solutions to Eqs. (\ref{udelta}) and (\ref{vdelta}):%
\begin{eqnarray}
u(x) &=&\frac{\left( 3k\right) ^{1/4}}{\sqrt{\cosh \left( \sqrt{8k}(|x|+\xi
\right) }},  \label{xi} \\
v(x) &=&\frac{\left( 3k\right) ^{1/4}}{\sqrt{\cosh \left( \sqrt{8k}(|x|+\eta
\right) }}.  \label{eta}
\end{eqnarray}%
Here, possible asymmetry between the two components of the soliton in the
symmetric system is represented by different positive offset parameters, $%
\xi \neq \eta $.

The effect of the delta-functional terms in Eqs. (\ref{udelta}) and (\ref%
{vdelta}) is represented by jumps (discontinuities) of the first derivatives
of $u(x)$ and $v(x)$ at point $x=0$, which are produced by the integration
of Eqs. (\ref{udelta}) and (\ref{vdelta}) in an infinitesimal vicinity of $%
x=0$, while functions $u(x)$ and $v(x)$ are continuous at this point:%
\begin{eqnarray}
\frac{du}{dx}(x &=&+0)-\frac{du}{dx}(x=-0)=-2\left[ v(x=0)+\varepsilon u(x=0)%
\right] ,  \label{jump1} \\
\frac{dv}{dx}(x &=&+0)-\frac{dv}{dx}(x=-0)=-2\left[ u(x=0)+\varepsilon v(x=0)%
\right] .  \label{jump2}
\end{eqnarray}%
The substitution of expressions (\ref{xi}) and (\ref{eta}) in Eqs. (\ref%
{jump1}) and (\ref{jump2}) leads to the following system of equations for $%
\xi $ and $\eta $:%
\begin{eqnarray}
2k\left( c_{1}^{2}-1\right) c_{2} &=&\left( c_{1}+\varepsilon
^{2}c_{2}+2\varepsilon \sqrt{c_{1}c_{2}}\right) c_{1}^{2},  \label{c1eps} \\
2k\left( c_{2}^{2}-1\right) c_{1} &=&\left( c_{2}+\varepsilon
^{2}c_{1}+2\varepsilon \sqrt{c_{1}c_{2}}\right) c_{2}^{2},  \label{c2eps}
\end{eqnarray}%
where
\begin{equation}
c_{1}\equiv \cosh \left( \sqrt{8k}\xi \right) ,c_{2}\equiv \cosh \left(
\sqrt{8k}\eta \right) .  \label{cc12}
\end{equation}

\subsection{Symmetric states}

First, for symmetric solutions, with $\xi =\eta $, i.e., $c_{1}=c_{2}$ in
Eq. (\ref{cc12}), the matching conditions given by Eqs. (\ref{c1eps}) and (%
\ref{c2eps}) coalesce into a single equation,%
\begin{equation}
\tanh \left( 2\sqrt{2k}\xi \right) =\frac{1+\varepsilon }{\sqrt{2k}},
\label{epstanh}
\end{equation}%
cf. Eq. (\ref{tanh}). The constraint $\tanh \left( 2\sqrt{2k}\xi \right) <1$
defines the existence range of the symmetric solitons,%
\begin{equation}
\frac{1}{2}\left( 1+\varepsilon \right) ^{2}<k<\infty ,  \label{eps1/2}
\end{equation}%
which is identical to range (\ref{1/2}) in the case of $\varepsilon =0$. The
power of the symmetric soliton can be easily obtained from Eqs. (\ref{xi}), (%
\ref{eta}), and (\ref{epstanh}):%
\begin{equation}
P_{\mathrm{symm}}(k)=\sqrt{6}\left[ \pi -2\arctan \left( \sqrt{\frac{\sqrt{2k%
}+(1+\varepsilon )}{\sqrt{2k}-(1+\varepsilon )}}\right) \right] .
\label{epsN}
\end{equation}%
This expression starts from $P_{\mathrm{symm}}=0$ at the left edge the
existence range (\ref{eps1/2}) and monotonously grows with the increase of $%
k $, up to $P_{\mathrm{symm}}(k\rightarrow \infty )=\sqrt{6}\pi /2\equiv 2P_{%
\mathrm{TS}}$, see Eq. (\ref{NTS}). Thus, similar to Eq. (\ref{N(mu)}, for
any value of $\varepsilon $ Eq. (\ref{epsN}) meets the VK criterion, $%
dP/dk>0 $. Furthermore, the symmetric solution given by Eqs. (\ref{xi}), (%
\ref{eta}), and (\ref{epstanh}) exists even for $-1<\varepsilon <0$, when
the local potential is moderately repulsive, as the effective attractive
potential induced by the fused coupling is stronger.

\subsection{Asymmetric states}

\subsubsection{The case of $\protect\varepsilon =0$}

It is easy to find a solution of Eqs. (\ref{c1eps}) and (\ref{c2eps}) with
broken symmetry, i.e., $c_{1}\neq c_{2}$, in the case of $\varepsilon =0$,
when the system does not include the delta-functional potential:

\begin{equation}
c_{1,2}^{2}=2k^{2}\pm 2k\sqrt{k^{2}-1}.  \label{c1,c2}
\end{equation}%
This solution exists, obviously, in the range of%
\begin{equation}
1<k<\infty .  \label{-1}
\end{equation}%
Note that condition $c_{1,2}>1$, which follows from definition (\ref{cc12}),
holds for solution (\ref{c1,c2}) in interval (\ref{-1}). Comparison with Eq.
(\ref{eps1/2}) demonstrates that, in the case of $\varepsilon =0$, only the
symmetric states exist in the narrow interval of $1/2<k<1$, while the
semi-infinite area (\ref{-1}) is populated by both the symmetric and
asymmetric modes.

The analytical calculation of the power for the asymmetric solution, which
is given, for $\varepsilon =0$, by Eqs. (\ref{xi}), (\ref{eta}), and (\ref%
{c1,c2}), yields a surprising exact result: it does not depend on $k$ [on
the contrary to power (\ref{epsN}) of the symmetric solitons in the same
case of $\varepsilon =0$], keeping the value given by Eq. (\ref{NTS}). It is
easy to check that, naturally, this single value of the norm is identical to
the norm of the symmetric solution at $k=1$, which is given by the doubled
expression (\ref{epsN}) with $k=1$ and $\varepsilon =0$. The fact that the
family of the asymmetric solitons is degenerate, featuring the single value
of the power, implies that it is another species of TS family, thus being
certainly unstable.

The asymmetry of the stationary states is characterized by the
power-imbalance parameter,%
\begin{equation}
\Theta \equiv \pm \frac{P_{u}-P_{v}}{P_{u}+P_{v}},  \label{theta}
\end{equation}%
where the opposite signs correspond to two branches of the asymmetric
states, that are mirror images of each other, see Figs. \ref{fig5}, \ref%
{fig6}, and \ref{fig13} below. The analytical solution (\ref{c1,c2})\ yields%
\begin{eqnarray}
\Theta (k,\varepsilon &=&0)=\pm \frac{4}{\pi }\left[ \arctan \left( \sqrt{%
2k^{2}+2k\sqrt{k^{2}-1}}+\sqrt{2k^{2}+2k\sqrt{k^{2}-1}-1}\right) \right.
\notag \\
&&\left. -\arctan \left( \sqrt{2k^{2}-2k\sqrt{k^{2}-1}}+\sqrt{2k^{2}-2k\sqrt{%
k^{2}-1}-1}\right) \right] .  \label{Theta}
\end{eqnarray}%
With the increase of $k$ from $1$ to $\infty $, as per Eq. (\ref{-1}), $%
\Theta $ given by Eq. (\ref{Theta}) monotonously varies from $0$ to $\pm 1$,
the latter limit implying that one core of the coupler gets asymptotically
empty. Usually, the SSB transition is characterized by dependence $\Theta
(P) $, rather than $\Theta (k)$, because the power and asymmetry are
observable parameters of optical beams, unlike the \textquotedblleft hidden"
propagation constant. However, in the present case it is irrelevant to refer
to $\Theta (P)$, as the degenerate family of the asymmetric solitons of the
TS type keeps the single value (\ref{NTS}) of the power at $\varepsilon =0$.

\subsubsection{The general case, $\protect\varepsilon \neq 0$}

In the case of $\varepsilon \neq 0$, equations (\ref{c1eps}) and (\ref{c2eps}%
) are too complex for the full analytical solution, Nevertheless, analysis
of these equations makes it possible to identify the point at which SSB
commences, i.e., a solution with an infinitesimal difference $c_{1}-c_{2}$
branching off from the symmetric one with $c_{1}=c_{2}\equiv c$. The
linearization of Eqs. (\ref{c1eps}) and (\ref{c2eps}) with respect to
infinitesimal $\left( c_{1}-c_{2}\right) $ demonstrates that this
bifurcation happens at
\begin{equation}
c=c_{\mathrm{SSB}}\equiv \sqrt{2+\varepsilon },~k\equiv k_{\mathrm{SSB}}=%
\frac{1}{2}\left( 1+\varepsilon \right) \left( 2+\varepsilon \right) ,
\label{SSB}
\end{equation}%
i.e., the asymmetric solitons exist in the range of%
\begin{equation}
\frac{1}{2}\left( 1+\varepsilon \right) \left( 2+\varepsilon \right)
<k<\infty .  \label{epsmu}
\end{equation}%
In the case of $\varepsilon =0$, this range is identical to one given above
by Eq. (\ref{-1}).

It is also instructive to compare values of the power at $k=k_{\mathrm{SSB}}$
and $k\rightarrow \infty $. It is easy to find
\begin{equation}
P\left( k=k_{\mathrm{SSB}}\right) =\sqrt{6}\left[ \pi -2\arctan \left( \sqrt{%
\varepsilon +2}+\sqrt{\varepsilon +1}\right) \right]  \label{smallest}
\end{equation}%
[in the case of $\varepsilon =0$, Eq. (\ref{smallest}) gives the value
identical to one given by Eq. (\ref{NTS})], while, in the limit of $%
k\rightarrow \infty ,$ the power is the same as given above by Eq. (\ref{NTS}%
). Then, an essential conclusion is that%
\begin{eqnarray}
P_{\mathrm{asymm}}\left( k=k_{\mathrm{SSB}}\right) &<&P_{\mathrm{asymm}%
}(k\rightarrow \infty )\text{,~at }\varepsilon >0,  \notag \\
P_{\mathrm{asymm}}\left( k=k_{\mathrm{SSB}}\right) &\geq &P_{\mathrm{asymm}%
}(k\rightarrow \infty )\text{,~at }\varepsilon \leq 0.  \label{conclusion}
\end{eqnarray}%
Actually, Eq. (\ref{conclusion}) implies that the VK criterion holds at $%
\varepsilon >0$, and does not hold at $\varepsilon \leq 0$. Therefore, the
asymmetric solitons may be stable in the former case (with the attractive
local potential), and are definitely unstable if the potential is repulsive.
It is shown below that a part of the family of the asymmetric solitons is
indeed stable for $\varepsilon >0$.

\section{Numerical results}

\subsection{The setup for the numerical analysis}

The exact solution for the asymmetric states, given by Eqs. (\ref{c1eps})
and (\ref{c2eps}), cannot be cast in a fully explicit form for $\varepsilon
\neq 0$, and the full stability analysis cannot be performed analytically
beyond the verification of the VK criterion. These problems should be
addressed by means of numerical methods. The numerical solutions and test of
their stability are produced below in the framework of the system in which
the ideal delta-function in Eqs. (\ref{eps}) and (\ref{eps2}) is replaced by
the usual Gaussian regularization,%
\begin{equation}
\tilde{\delta}(x)=\left( \sqrt{\pi }\sigma \right) ^{-1}\exp \left( -\frac{%
x^{2}}{\sigma ^{2}}\right) ,  \label{Gaussian}
\end{equation}%
with small width $\sigma $. We here fix $\sigma =0.02$, which is
sufficiently small in comparison with the transverse width of various
trapped modes. Comparison between the exact analytical solutions produced
above for the system with the ideal delta-function and numerical results
obtained with the help of regularization (\ref{Gaussian}) is demonstrated
below in Figs. \ref{fig7} and \ref{fig8}. Note also that the system with the
ideal delta-function replaced by regularization (\ref{Gaussian}) is
appropriate for modeling real fused couplers, with a finite size of the
coupling region.

Numerical solution of Eqs. (\ref{udelta}) and (\ref{vdelta}) with $\delta
(x) $ replaced by $\tilde{\delta}(x)$ was performed by means of the
Newton-conjugate-gradient method, which readily provides robust convergence
to stationary solutions of the system \cite{Newton,Yang}. Stability of the
stationary solutions against small perturbations, taken in the usual form
\cite{Yang},
\begin{eqnarray}
\psi _{1}(x,z) &=&\exp (ikz)\left[ \exp (\gamma z)u_{1}(x)+u_{2}^{\ast
}(x)\exp (\gamma ^{\ast }z)\right] ,  \notag \\
\phi _{1}(x,z) &=&\exp (ikz)\left[ \exp (\gamma z)v_{1}(x)+v_{2}^{\ast
}(x)\exp (\gamma ^{\ast }z)\right] ,  \label{pert}
\end{eqnarray}%
is determined by the linearized %Bogoliubov -- de Gennes
equations for perturbation amplitudes $u_{1,2}(x)$ and $v_{1,2}(x)$ and
instability eigenvalue (growth rate) $\gamma $:%
\begin{eqnarray}
-(k-i\gamma )u_{1} &=&-\frac{1}{2}\frac{d^{2}u_{1}}{dx^{2}}%
-3u^{4}(x)u_{1}-2u^{4}(x)u_{2}-\delta (x)\left( v_{1}+\varepsilon
u_{1}\right) ,  \notag \\
-(k+i\gamma )u_{2} &=&-\frac{1}{2}\frac{d^{2}u_{2}}{dx^{2}}%
-3u^{4}(x)u_{2}-2u^{4}(x)u_{1}-\delta (x)\left( v_{2}+\varepsilon
u_{2}\right) ,  \notag \\
-(k-i\gamma )v_{1} &=&-\frac{1}{2}\frac{d^{2}v_{1}}{dx^{2}}%
-3v^{4}(x)v_{1}-2v^{4}(x)v_{2}-\delta (x)\left( u_{1}+\varepsilon
v_{1}\right) ,  \notag \\
-(k+i\gamma )v_{2} &=&-\frac{1}{2}\frac{d^{2}v_{2}}{dx^{2}}%
-3v^{4}(x)v_{2}-2v^{4}(x)v_{1}-\delta (x)\left( u_{2}+\varepsilon
v_{2}\right) .  \label{BdG}
\end{eqnarray}%
The ordinary stability condition is that all eigenvalues $\gamma $ produced
by Eqs. (\ref{BdG}) must have zero real parts. The spectrum of the
eigenvalues was produced by means of the Fourier collocation method.

Predictions for the (in)stability of the stationary states, provided by the
numerical computation of the eigenvalues for small perturbations, were
verified by means of simulations of the perturbed evolution in the framework
of Eqs. (\ref{eps1}) and (\ref{eps2}), again with $\delta (x)$ replaced by $%
\tilde{\delta}(x)$. Characteristic examples of stable and unstable evolution
are produced below.

\subsection{Stable and unstable families of symmetric and asymmetric bound
states}

\begin{figure}[tbp]
\begin{center}
\includegraphics[width=0.80\textwidth]{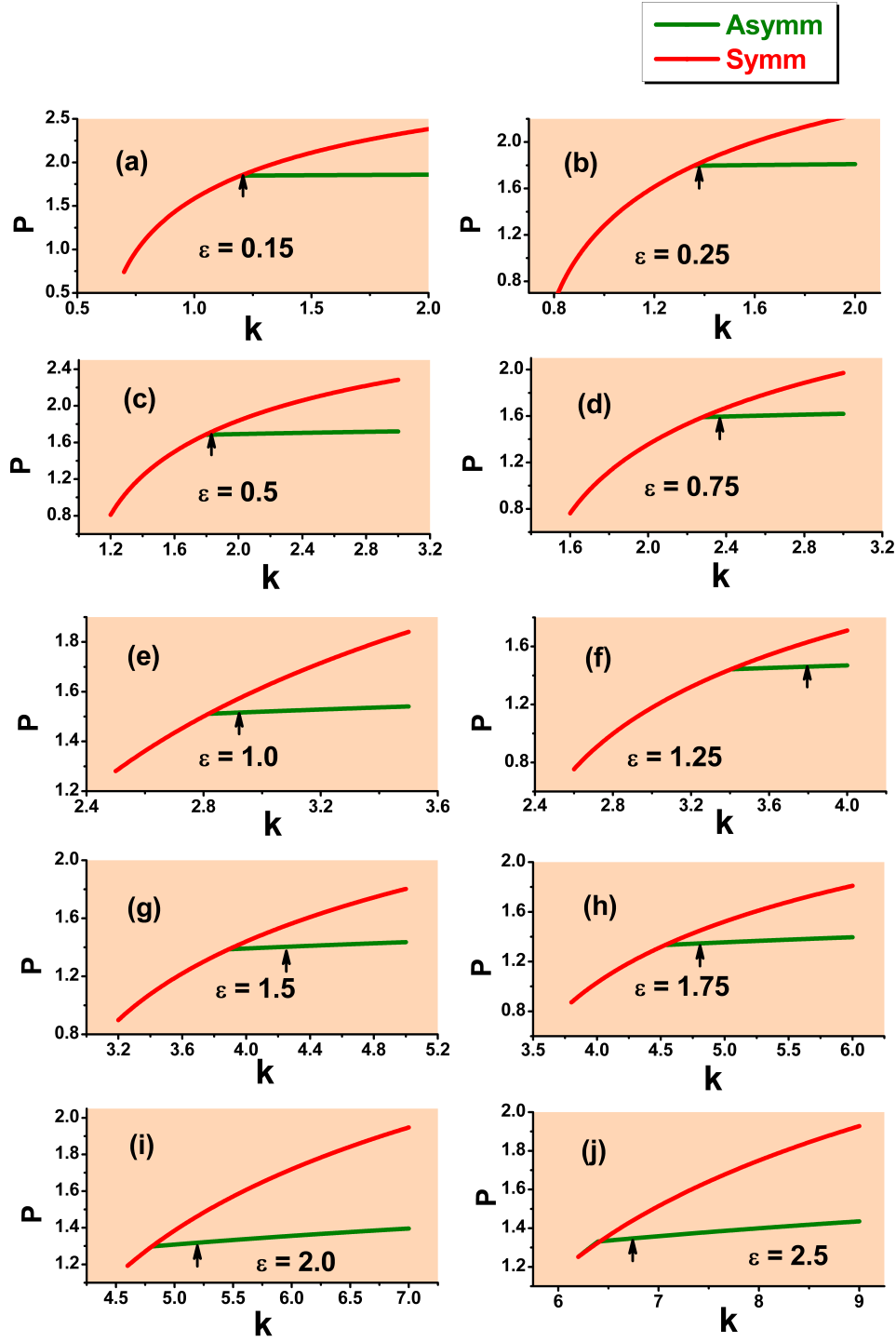}
\end{center}
\caption{Power $P$ of symmetric and asymmetric bound states (solitons) vs.
the propagation constant $k$ for different values of strength $\protect%
\varepsilon $ of the attractive potential in Eqs. (\protect\ref{eps1}) and (%
\protect\ref{eps2}), which are indicated in panels. Note essentially
different scales of $k$ in different panels. Families of asymmetric states
are represented by nearly horizontal branches originating at SSB
(spontaneous-symmetry-breaking) bifurcation points. The results are produced
by the numerical solution of Eqs. (\protect\ref{udelta}) and (\protect\ref%
{vdelta}), where the ideal delta-function is replaced by the narrow Gaussian
(\protect\ref{Gaussian}) with $\protect\sigma =0.02$. The families of
symmetric states are stable and unstable before and after the bifurcation
(to the left and right of the SSB points, respectively). The branches of the
asymmetric states are stable in segments between the SSB points and
stability boundaries designated by arrows. The stability segments are
extremely narrow for $\protect\varepsilon \leq 0.5$, in which cases the
arrows are located very close to the SSB points. The stability is identified
according to eigenvalues produced by the numerical solution of Eqs. (\protect
\ref{BdG}).}
\label{fig1}
\end{figure}

\begin{figure}[tbp]
\begin{center}
\includegraphics[width=0.80\textwidth]{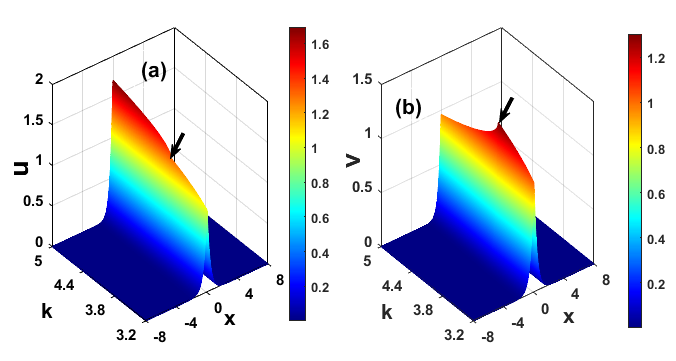}
\end{center}
\caption{The evolution of profiles $u(x)$ and $v(x)$ of two components of
the bound states [panels (a) and (b), respectively] following the increase
of $k$, at fixed value $\protect\varepsilon =1.5$ of the attractive
potential. The SSB point, $k_{\mathrm{SSB}}\approx 3.85$, is indicated by
black arrows. The profiles are produced by the numerical solution of Eqs. (%
\protect\ref{udelta}) and (\protect\ref{vdelta}), where $\protect\delta (x)$
is replaced by Gaussian (\protect\ref{Gaussian}) with $\protect\sigma =0.02$%
. Profiles of the symmetric state are not plotted at $k>k_{\mathrm{SSB}}$,
where they are definitely unstable. According to Figs. \protect\ref{fig1}(g)
and \protect\ref{fig4}(b) (see below), the asymmetric states are stable in
the interval of size $\Delta k$ $\approx 0.4$ between $k=k_{\mathrm{SSB}}$
and $k=k_{\mathrm{SSB}}+\Delta k$.}
\label{fig2}
\end{figure}

First, results of the numerical solution of Eqs. (\ref{udelta}) and (\ref%
{vdelta}), with $\delta (x)$ replaced by regularization (\ref{Gaussian}),
are collected, in the form of dependences $P(k)$ for the families of
symmetric and asymmetric bound states, in Fig. \ref{fig1}. The plots also
present conclusions for the stability, as obtained from the numerical
solution of Eqs. (\ref{BdG}). Typical examples of the symmetric and
asymmetric solitons, both stable and unstable ones, are displayed in Fig. %
\ref{fig2}.

These results are reported for the strength of the attractive potential
varying in the interval of
\begin{equation}
0.15\leq \varepsilon \leq 2.5.  \label{eps}
\end{equation}%
At $\varepsilon <0.15$, the findings are virtually identical to those for $%
\varepsilon =0.15$. The case of $\varepsilon >2.5$ is not presented here, as
in that case the strong attractive potential makes the bound state so narrow
that its width is no longer much larger than the fixed width $\sigma =0.02$
of the regularized delta-function, adopted in Eq. (\ref{Gaussian}); for this
reason, the setup becomes essentially different from the one defined by Eqs.
(\ref{eps1}) and \ref{eps2}). This is shown, in particular, below in Fig. %
\ref{fig8}.

Shapes of the symmetric and asymmetric solitons, and their evolution
following the increase of the propagation constant, $k$, are displayed in
Fig. \ref{fig2}. The figure clearly shows the change in the shape triggered
by the onset of SSB at the point which is indicated by arrows.

Naturally, symmetric solitons are stable below the SSB\ point, and unstable
above it. As shown above by means of the VK criterion, the asymmetric
solitons are completely unstable at $\varepsilon =0$. As an extension of
that finding, Fig. \ref{fig1} shows that the asymmetric states remain almost
completely unstable at $\varepsilon \leq 0.5$, while a visible stability
interval, of width $\Delta k$, appears above the SSB point at $\varepsilon
>0.5$. %This conclusion is illustrated in more detail by Fig. \ref%
%{fig3}, which displays the instability growth rate, i.e., Re$(\gamma )>0$
%[see Eq. (\ref{BdG})], vs. $k$. Actually, the unstable eigenvalues for the
%asymmetric solitons are complex, although Im$(\gamma )$ is not displayed
%here.
%\begin{figure}[tbp]
%\begin{center}
%\includegraphics[width=0.90\textwidth]{Fig3.png}
%\end{center}
%\caption{The instability growth rate of the symmetric and asymmetric bound
%states, Re$(\protect\gamma )$, vs. $k$ for different values of $\protect%
%\varepsilon $ which are indicated in panels. The results are produced by the
%numerical solution of Eqs. (\protect\ref{BdG}), where the ideal
%delta-function is replaced by the narrow Gaussian (\protect\ref{Gaussian})
%with $\protect\sigma =0.02$. The bound states are stable in regions where Re$%
%(\protect\gamma )=0$ [zero values of Re$(\protect\gamma )$ are not
%explicitly shown]. The curve Re$(\protect\gamma )$ for the symmetric
%solitons commences at the SSB point. The asymmetric solitons are stable in
%bold segments (referred to as $\Delta k$ in the text) of axis $k$.}
%\label{fig3}
%\end{figure}

\begin{figure}[tbp]
\begin{center}
\includegraphics[width=0.50\textwidth]{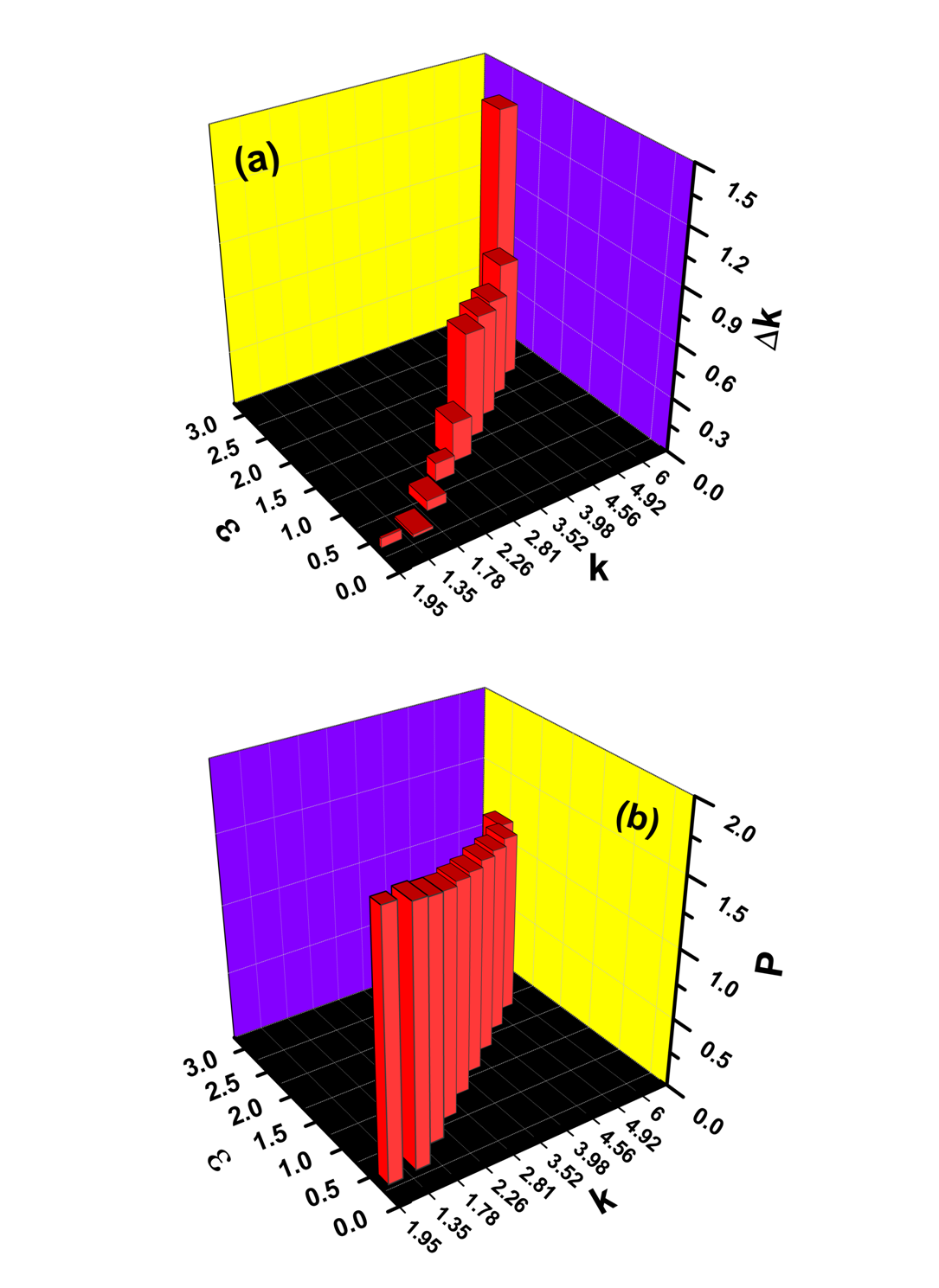}
\end{center}
\caption{(a) Width $\Delta k$ of the stability interval for stable
asymmetric bound states vs. strength $\protect\varepsilon $ of the
attractive potential in Eqs. (\protect\ref{eps1}) and (\protect\ref{eps2}).
The corresponding values $k$ in the plane of $\left( \protect\varepsilon %
,k\right) $ indicate the midpoint of the stability interval. (b) Values of
power $P$ at the midpoint.}
\label{fig4}
\end{figure}

The dependence of width $\Delta k$ of the stability interval for the
asymmetric bound states on $\varepsilon $ is shown in Fig. \ref{fig4}(a). In
addition, panel (b) of Fig. \ref{fig4} shows values of the power for the
asymmetric bound states at the midpoint of the respective stability
intervals. These results, produced by the computation of stability
eigenvalues, are confirmed by direct simulations of the perturbed evolution
of the asymmetric bound states, as shown below in Figs. \ref{fig9}-\ref%
{fig12}.

\begin{figure}[tbp]
\begin{center}
\includegraphics[width=0.60\textwidth]{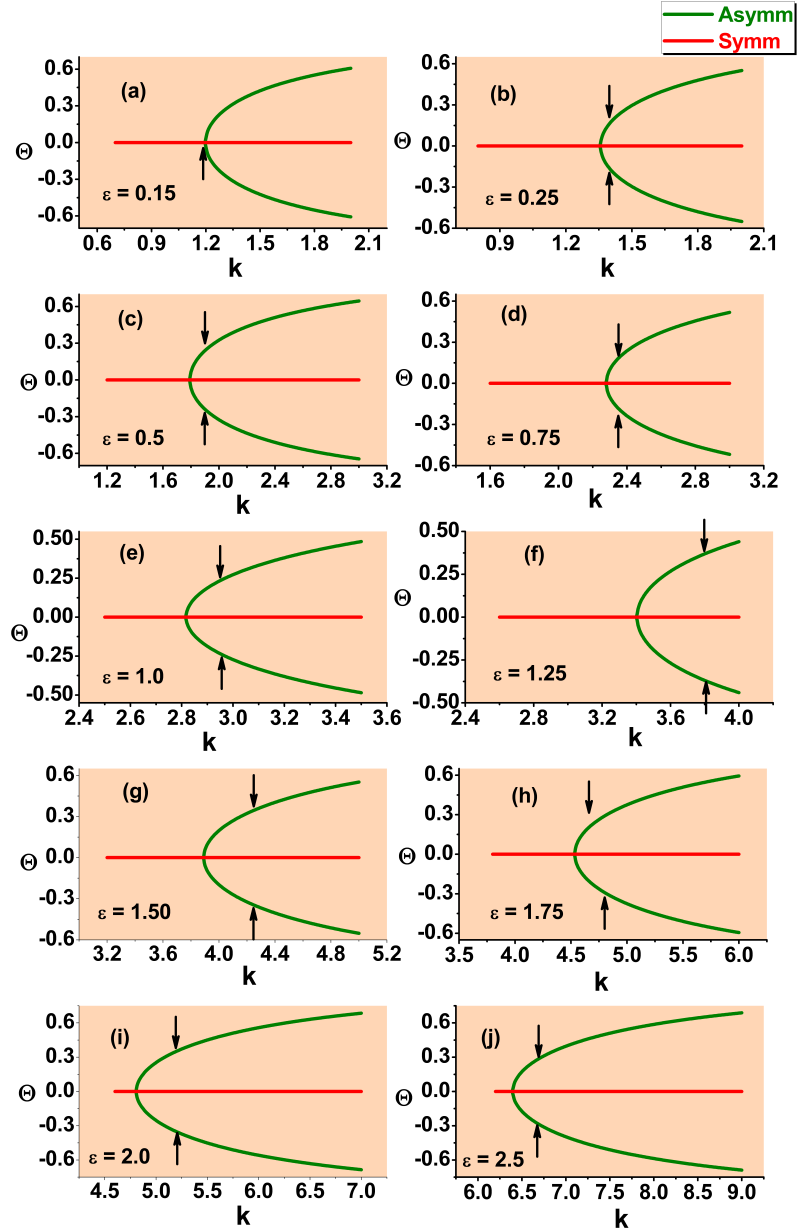}
\end{center}
\caption{The SSB bifurcation represented by the dependence of the
numerically calculated values of the asymmetry parameter $\Theta $ for the
asymmetric bound states [see Eq. (\protect\ref{theta})] on propagation
constant $k$. $\Theta =0$ corresponds to the symmetric bound states, which
are stable to the left of the SSB point, and unstable to the right of it.
The branches of the symmetry-broken states are stable in intervals between
the SSB\ point and stability boundary designated by arrows. For $\protect%
\varepsilon =0.15$ only one arrow is shown, as its location practically
coincides with the SSB point. Note the difference of the scale of $k$ in
different panels.}
\label{fig5}
\end{figure}

\begin{figure}[tbp]
\begin{center}
\includegraphics[width=0.60\textwidth]{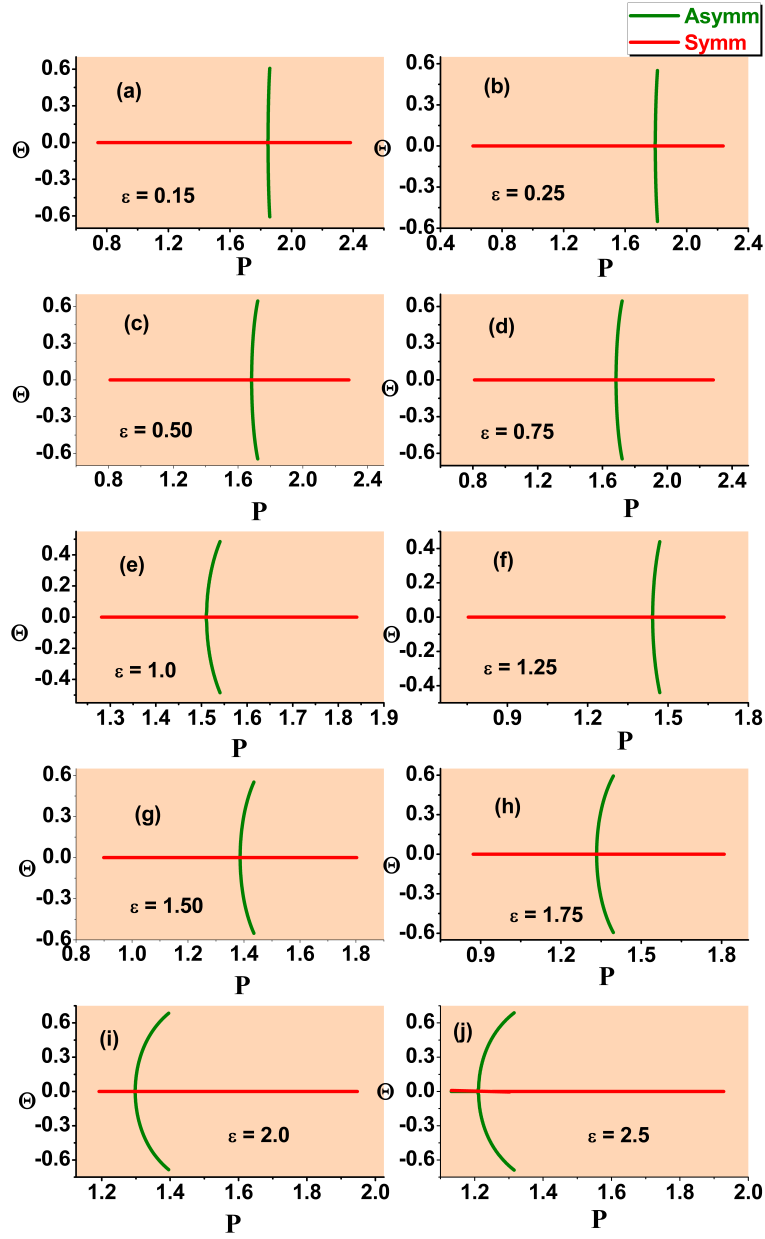}
\end{center}
\caption{The SSB bifurcation represented by dependences $\Theta (P)$ of the
asymmetry parameter on the total power. The branches of the symmetry-broken
states are not extended very close to limit values $\Theta =\pm 1$, which
correspond to $k\rightarrow \infty $, as the respective width of the bound
states, $W\simeq k^{-1/2}$ [see Eqs. (\protect\ref{xi}) and (\protect\ref%
{eta})], ceases to be much larger than width $\protect\sigma $ of the
regularized delta-function (\protect\ref{Gaussian}).}
\label{fig6}
\end{figure}

It is relevant to present the asymmetric-soliton families in terms of the
asymmetry parameter $\Theta $, defined as per (\ref{theta}). To this end,
dependences $\Theta (k)$ and $\Theta (P)$ are plotted in Figs. \ref{fig5}
and \ref{fig6}, respectively. The virtually flat vertical shape of the $%
\Theta (P)$ lines for $\varepsilon <0.5\ $in Fig. \ref{fig6} corresponds to
the above analytical result obtained for $\varepsilon =0$, according to
which the power of the whole family of the asymmetric solitons takes the
single value (\ref{NTS}). It is relevant to stress that the $\Theta (P)$
curves which are not fully flat feature the forward-going (convex) shape,
clearly indicating that the SSB bifurcation in the present system is of the
supercritical type \cite{bifurcations}, i.e., it represents the phase
transition of the second kind.

\subsection{Comparison of the numerical and analytical results}

\begin{figure}[tbp]
\begin{center}
\includegraphics[width=0.60\textwidth]{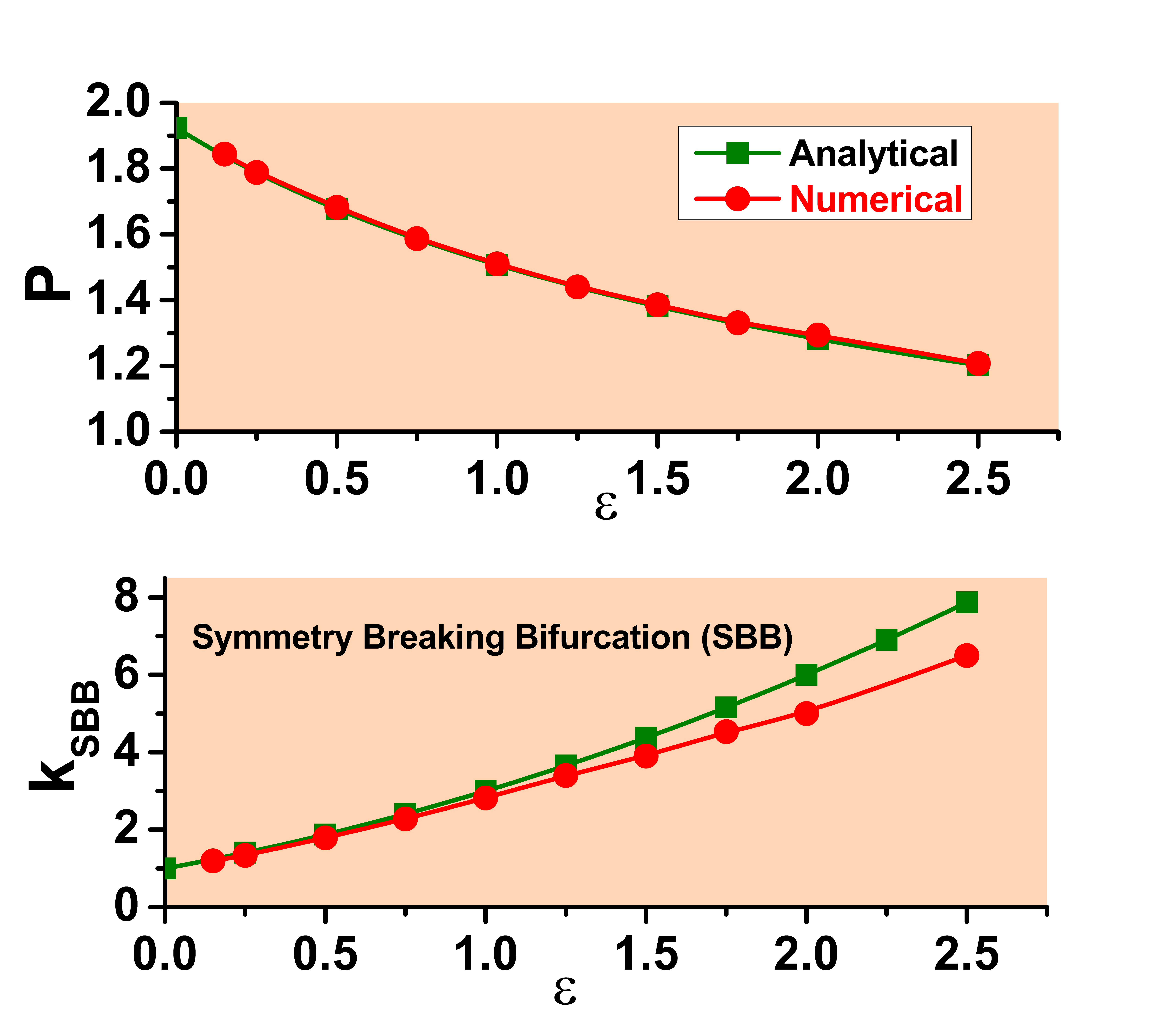}
\end{center}
\caption{The comparison of the analytical prediction for the critical values
of the power and propagation constant (the top and bottom panels,
respectively) at the SSB point, as given by Eq. (\protect\ref{PK}), and
their counterparts produced by the numerical solution of Eqs. (\protect\ref%
{udelta}) and (\protect\ref{vdelta}), in which the delta-function\ is
replaced by the regularized expression (\protect\ref{Gaussian}) with $%
\protect\sigma =0.02$.}
\label{fig7}
\end{figure}

It is relevant to compare the analytical predictions, obtained above in the
model based on Eqs. (\ref{eps1}) and (\ref{eps2}) with the ideal
delta-function, and systematic numerical results produced for the system
with the delta-function replaced by the regularized expression (\ref%
{Gaussian}). First, in Fig. \ref{fig7} we compare the analytical values of
the power and propagation constant at the SSB point, as given by Eq. (\ref%
{PK}), and their numerically found counterparts, as functions of $%
\varepsilon $ in the interval (\ref{eps}) under the consideration. The top
panel of Fig. \ref{fig7} clearly demonstrates that the analytically
predicted critical values of $P$ are virtually identical to the numerical
ones. Further, the bottom panel shows that the analytical and numerical
critical values of $k$ are almost identical at $\varepsilon <1$, and a
relatively small discrepancy appears at $\varepsilon >1$. As mentioned
above, it is explained by the fact that strong attraction to the central
point by the local potential makes the trapped state so narrow that its
width ceases to be much larger than the finite size $\sigma $ of the
regularized delta-functional profile in Eq. (\ref{Gaussian}).

\begin{figure}[tbp]
\begin{center}
\includegraphics[width=0.50\textwidth]{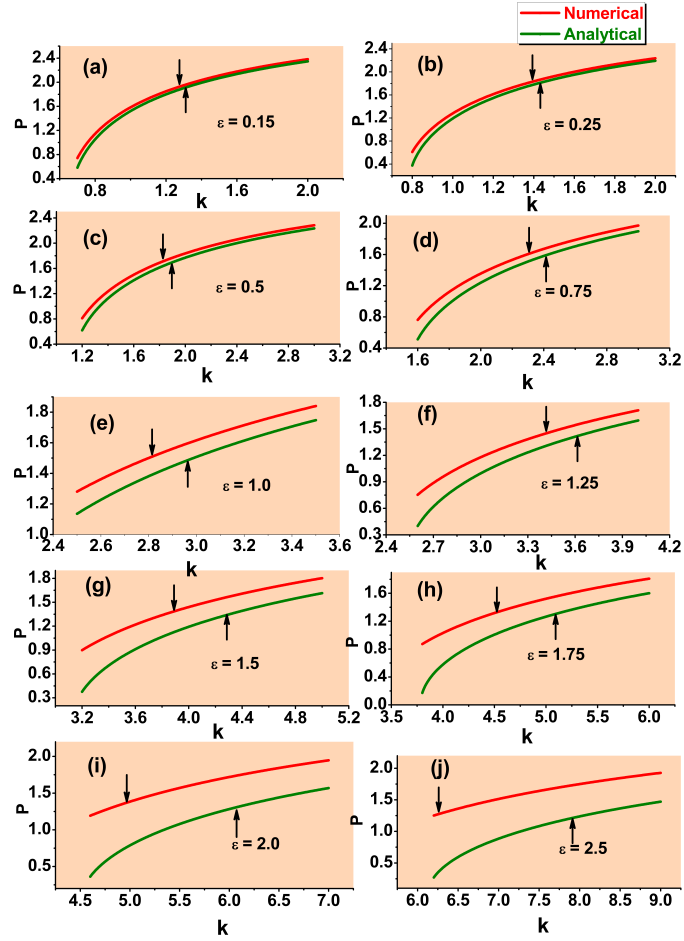}
\end{center}
\caption{The comparison of the $P(k)$ dependence for the family of symmetric
bound states, produced in the analytical form by Eq. (\protect\ref{epsN}),
and its counterpart, produced by the numerical solution of Eqs. (\protect\ref%
{udelta}) and (\protect\ref{vdelta}) with $\protect\delta (x)$ replaced by
expression (\protect\ref{Gaussian}), in which $\protect\sigma =0.02$ is
fixed. Arrows demonstrate the location of the SSB points on the numerical
and analytical curves, the symmetric states being stable before the
bifurcation, i.e., to the left of the indicated points.}
\label{fig8}
\end{figure}

Further comparison of the analytical and numerical results is provided by
Fig. \ref{fig8}, in terms of the $P(k)$ curves for the families of symmetric
solitons (the numerical curves are essentially the same as shown, also for
the symmetric states, in Fig. \ref{fig1}). This picture again demonstrates
close proximity of the analytical and numerical results at $\varepsilon <1$,
and gradual increase of the discrepancy at $\varepsilon >1$.

\begin{figure}[tbp]
\begin{center}
\includegraphics[width=0.60\textwidth]{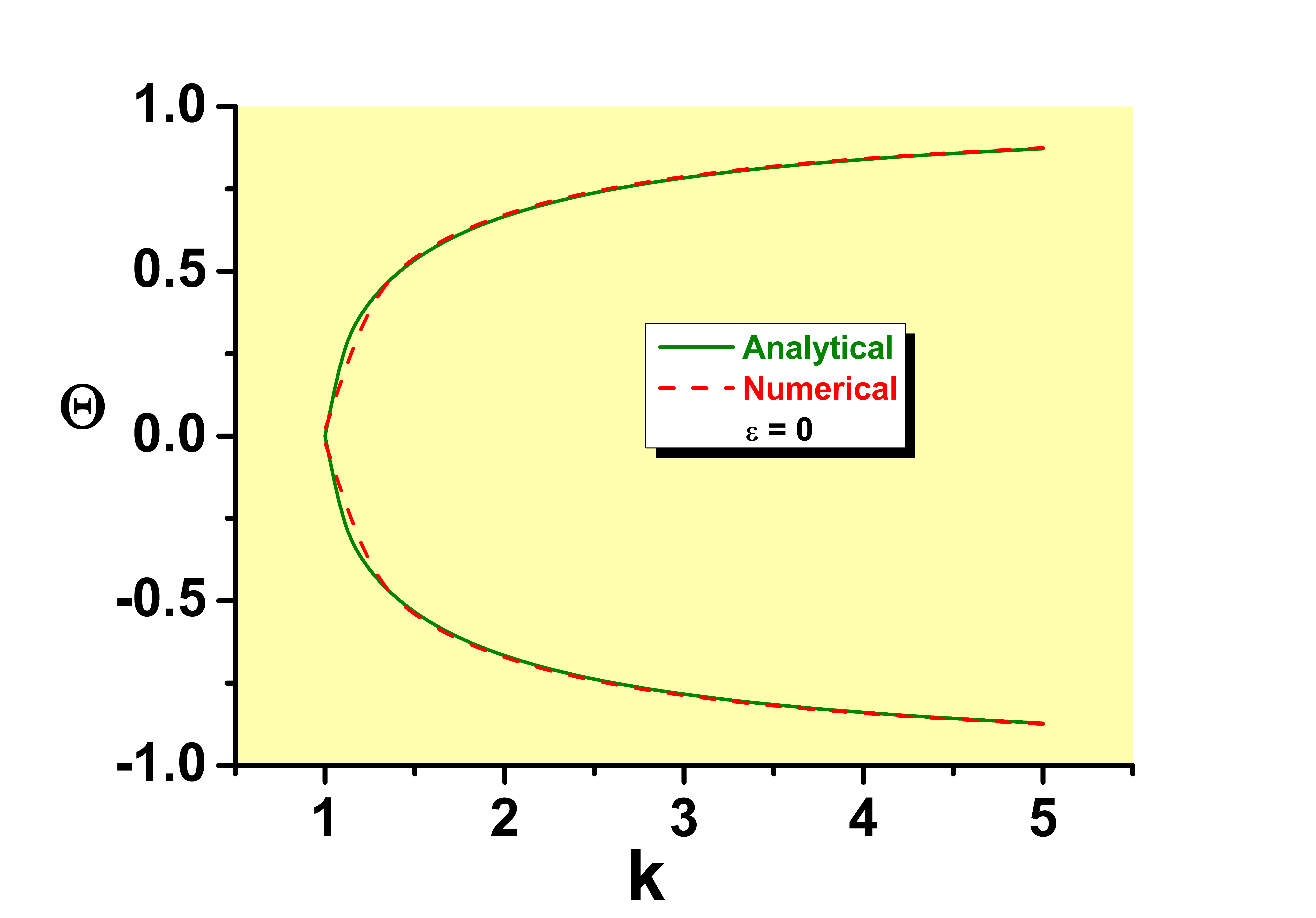}
\end{center}
\caption{The comparison of the analytical $\Theta (k)$ dependence for $%
\protect\varepsilon =0$, as given by Eq. (\protect\ref{Theta}), and its
numerically generated counterpart.}
\label{fig13}
\end{figure}

It is also natural to compare the analytical dependence of the asymmetry
parameter, $\Theta (k)$, given for $\varepsilon =0$ by Eq. (\ref{Theta}),
and its numerically generated counterpart, which is shown in Fig. \ref{fig13}%
. It is seen that both dependences match perfectly, being also very close to
the numerically generated $\Theta (k)$ dependence plotted for $\varepsilon
=0.15$ in Fig. \ref{fig5}(a).

\subsection{Perturbed evolution of stable and unstable bound states}

\begin{figure}[tbp]
\begin{center}
\includegraphics[width=0.90\textwidth]{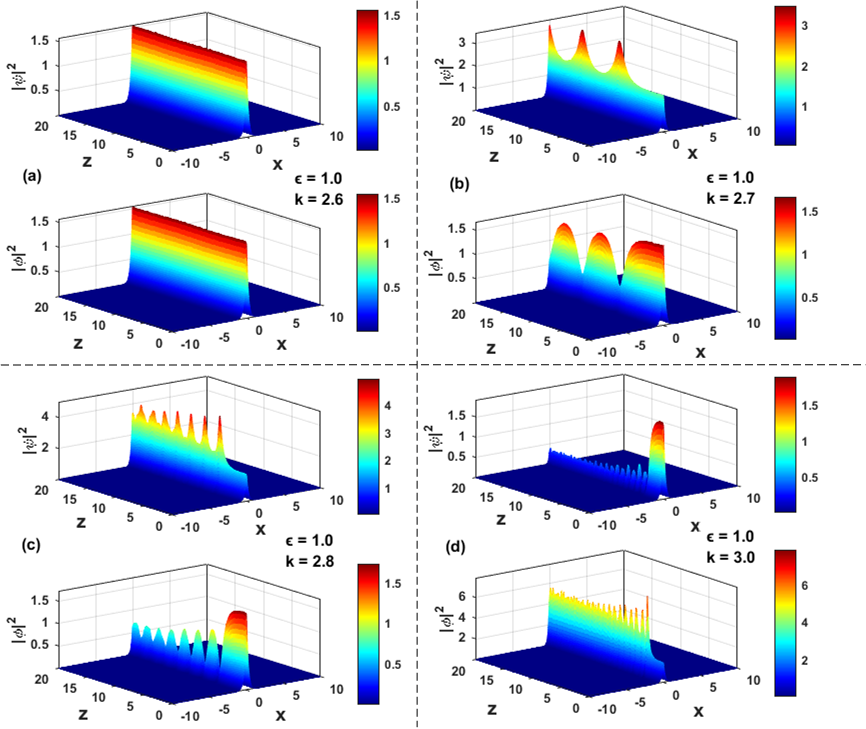}
\end{center}
\caption{Direct simulations of the evolution of the initially symmetric
bound states at $\protect\varepsilon =1$ and values of the propagation
constant $k$ indicated in the panels. As expected, the symmetric state is
stable below the SSB point [in panel (a)], and unstable above it. Note
different scales on vertical axes in the plots of components $\protect\psi $
and $\protect\phi $.}
\label{fig9}
\end{figure}

\begin{figure}[tbp]
\begin{center}
\includegraphics[width=0.90\textwidth]{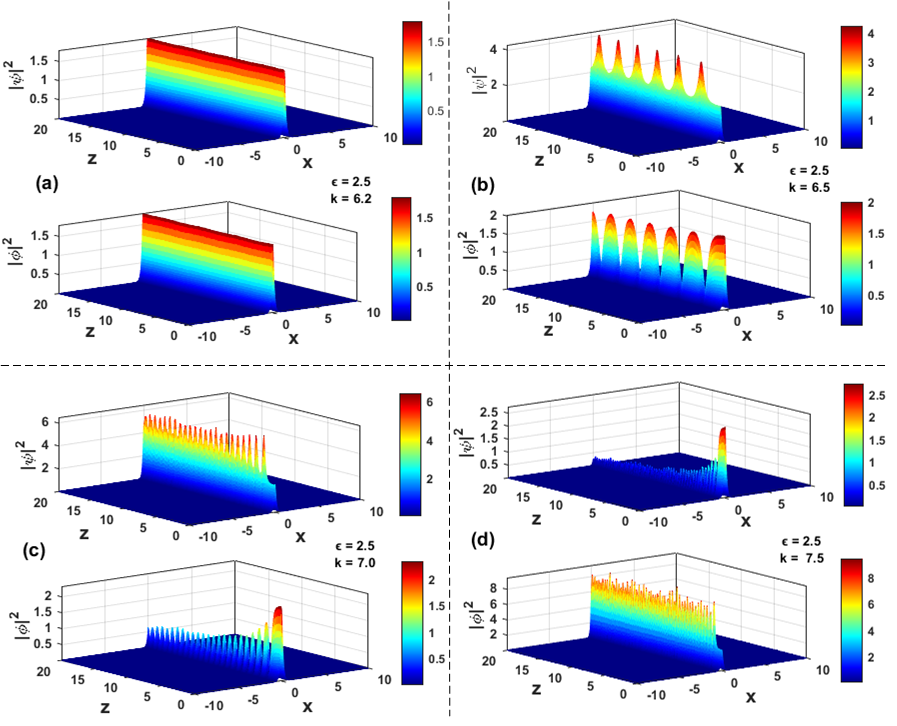}
\end{center}
\caption{The same as in Fig. \protect\ref{fig9}, but for the largest
strength of the attractive potential considered here, i.e., $\protect%
\varepsilon =2.5$.}
\label{fig10}
\end{figure}

The predictions for the (in)stability, which are produced above in the
analytical form by means of the VK criterion and by dint of the numerical
solution of the eigenvalue problem based on Eq. (\ref{BdG}), have been
verified by direct simulations of Eqs. (\ref{eps1}) and (\ref{eps2}), with $%
\delta (x)$ again replaced by approximation (\ref{Gaussian}) with $\sigma
=0.02$. First, the simulations of the symmetric solitons, which are reported
in Figs. \ref{fig9} and \ref{fig10} for weaker and stronger attractive
potentials, \textit{viz}., $\varepsilon =1$ and $2.5$, demonstrate, as
expected, that these states remain stable at $k<k_{\mathrm{SSB}}$, and
develop spontaneous instability at $k>k_{\mathrm{SSB}}$. A clear trend is
that the instability makes amplitudes of the originally symmetric components
different, and initiates Josephson oscillations between them. In accordance
with the fact that the solution of Eq. (\ref{BdG}) produces purely real
instability eigenvalues $\gamma $ for the symmetric solitons at $k>k_{%
\mathrm{SSB}}$, the oscillation frequency is small close to the SSB point
[e.g., at $k=2.7$ in Fig. \ref{fig9}(b)]. As a result, the system
establishes robust breathers with different amplitudes in their two
components, as seen in Figs. \ref{fig9}(b,c) and \ref{fig10}(b,c).

Farther from the bifurcation (at larger values of $k$), the oscillations are
much faster, and the system quickly develops a state with a large difference
in amplitudes of the two components [e.g., at $k=3$ in Fig. \ref{fig9}(d),
and at $k=7.5$ in Fig. \ref{fig10}(d)]. The latter states seem as robust
strongly asymmetric ones with irregular small-amplitude intrinsic
vibrations. They approximately resemble stationary bound states with strong
asymmetry between the components; however, strictly stationary asymmetric
states do not exist for the same values of parameters. For instance, in the
two above-mentioned cases, corresponding to $k=3$ in Fig. \ref{fig9}(d) and $%
k=7.5$ in Fig. \ref{fig10}(d), the respective value of the power of the
symmetric solitons is $P\approx 1.6$ in either case, as seen from Figs. \ref%
{fig1}(e) and (j), respectively; on the other hand, the same figures
demonstrate that stationary asymmetric states do not exist for this $P$ --
for instance, the largest available power of the asymmetric solitons is $%
\approx 1.5$ in the former case, and only $\approx 1.3$ in the latter one.
Thus, the conclusion is that robust modes with strong asymmetry and large
values of the power are admitted by the system, but in the weakly vibrating
state.

\begin{figure}[tbp]
\begin{center}
\includegraphics[width=0.90\textwidth]{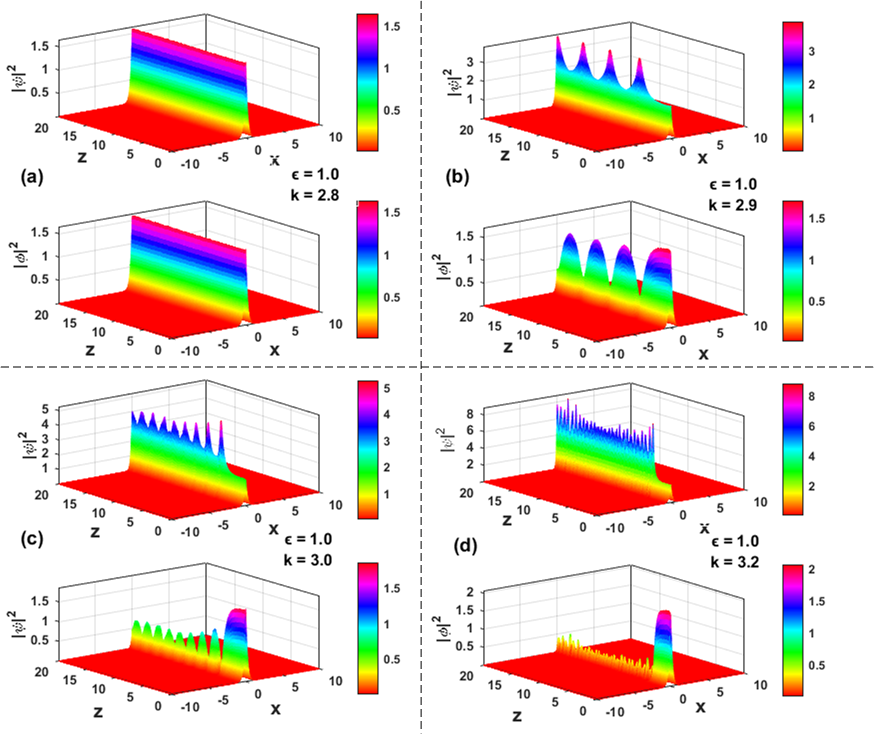}
\end{center}
\caption{The same as in Fig. \protect\ref{fig9} (with $\protect\varepsilon %
=1 $), but for the evolution of initially asymmetric bound states, which are
stable in (a) and unstable in (c-d).}
\label{fig11}
\end{figure}

\begin{figure}[tbp]
\begin{center}
\includegraphics[width=0.90\textwidth]{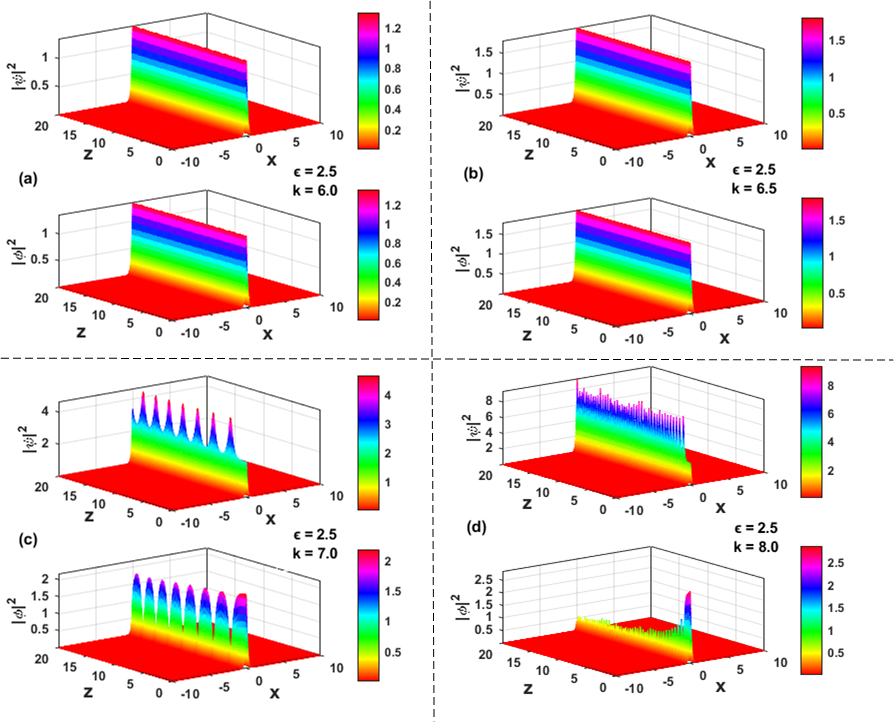}
\end{center}
\caption{The same as in Fig. \protect\ref{fig11} but for $\protect%
\varepsilon =2.5$.}
\label{fig12}
\end{figure}

Simulations of the evolution of asymmetric states corroborate that these
solitons are also stable or unstable in the cases when the numerical
solution of Eq. (\ref{BdG}) produces, respectively, zero or nonzero values
of the instability growth rate, Re$(\gamma )$. Typical examples for the same
values, $\varepsilon =1$ and $2.5$, as used in Figs. \ref{fig9} and \ref%
{fig10}, are displayed, severally, in Figs. \ref{fig11} and \ref{fig12}. It
is observed that the instability emerges in an oscillatory form, with a
finite frequency, in accordance with the fact that the instability
eigenvalues for the asymmetric states are found with nonzero imaginary
parts, unlike the above-mentioned pure real unstable eigenvalues for the
symmetric bound states.

In the case of relatively weak instability, the stationary asymmetric states
are spontaneously transformed into apparently robust asymmetric breathers,
which exhibit regular oscillations in Figs. \ref{fig11}(b,c) and \ref{fig12}%
(c). They are similar to the above-mentioned asymmetric breathers created by
the instability of the symmetric states, as shown in Figs. \ref{fig9}(b,c)
and \ref{fig10}(b,c).

On the other hand, strong instability, as seen in Figs. \ref{fig11}(d) and %
\ref{fig12}(d), creates states with a very strong asymmetry and
small-amplitude irregular intrinsic vibrations. The robust trapped states of
the latter type are similar to the above-mentioned strongly asymmetric modes
with irregular internal oscillations which are observed in Figs. \ref{fig9}%
(d) and \ref{fig10}(d)

\section{Conclusion}

This work aims to introduce a physically relevant model which admits the
exact solution for two basic problems, \textit{viz}., the stabilization of
TSs (Townes solitons) and SSB (spontaneous symmetry breaking) in nonlinear
couplers. The interest to these problems is drawn, in particular, by the
recently reported first experimental demonstration of weakly unstable 2D TSs
in BEC \cite{TS-exper,TS-exper 2} and SSB of optical solitons in dual-core
fibers \cite{Bugar}. The present model represents the system of parallel 1D
waveguides with the intrinsic quintic nonlinearity, which is critical in the
1D case, giving rise to the corresponding TSs. The waveguides form a fused
coupler, with the linear connection introduced in a narrow region. In
addition to that, the region of the local coupling carries the attractive
potential acting in each waveguiding core. The system can be directly
realized in optics, and it may be used in the design of photonic devices
operating in coupling and switching regimes. %In addition to that, it may
%serve as an approximate model of the super-Tonks-Girardeau gas of ultracold
%atoms loaded in the fused double cigar-shaped trap.
The model admits full analytical solutions for symmetric and asymmetric
solitons, revealing the explicit picture of the SSB in solitons and a
straightforward scenario for the stabilization of the TSs, which are
completely unstable in the uniform waveguides. It is found that the SSB
bifurcation of the supercritical type, i.e., the phase transition of the
second kind, destabilizes the symmetric bound states and gives rise to the
asymmetric ones, which remain stable in a finite interval of values of the
propagation constant. The evolution of those symmetric and asymmetric states
which are unstable leads to robust moderately asymmetric breathers, which
perform Josephson oscillations in the coupler, or (also robust) strongly
asymmetric states with small-amplitude irregular internal vibrations.

It is relevant to estimate parameters of the fused optical coupler which can
realize the proposed setup. An appropriate material is the above-mentioned
colloidal suspension of silver nanoparticles, with the volume-filling factor
$\simeq 1.5\times 10^{-5}$ and the corresponding quintic susceptibility $%
\chi ^{(5)}\simeq $ $4\times 10^{-38}$ m$^{4}/$V$^{4}$ at a visible
wavelength $\simeq 0.5$ $\mathrm{\mu }$m, while the cubic nonlinearity is
negligible \cite{Cid-OptExp,Cid}. Appropriate thickness of each waveguide,
as well as the width of the bridge connecting them, is $\simeq 3$ $\mathrm{%
\mu }$m. Then, the predicted solitons, with the transverse size $\gtrsim 30$
$\mathrm{\mu }$m, can be created by laser beams with power density about $10$
GW/cm$^{2}$ and total power $\sim 10$ kW. The propagation distance $\gtrsim 3
$ cm is sufficient to create well-formed solitons.

As an extension of the analysis, it may be relevant to introduce a system
with a double fused coupler, similar to the one considered in Ref. \cite%
{Alon}. Such a system should make it possible to study double SSB effects --
between the two cores of the coupler and, in the spatial direction, between
two separated regions of the fused coupling.

\section*{Acknowledgment}

This work was supported, in part, by the Israel Science Foundation through
grant No. 1695/22.


\begin{thebibliography}{99}
\bibitem{Berge} L. Berg\'{e}, Wave collapse in physics: principles and
applications to light and plasma waves, Phys. Rep. \textbf{303}, 259-370
(1998).

\bibitem{Sulem} C. Sulem and P.-L. Sulem, \textit{The Nonlinear Schr\"{o}%
dinger Equation: Self-Focusing and Wave Collapse} (Springer, New York, 1999).

\bibitem{Zakh-Kuz} V. E. Zakharov V. E. and E. A. Kuznetsov, Solitons and
collapses: two evolution scenarios of nonlinear wave systems, Physics --
Uspekhi \textbf{55}, 535-556 (2012).

\bibitem{Fibich} G. Fibich, \textit{The Nonlinear Schr\"{o}dinger Equation:
Singular Solutions and Optical Collapse} (Springer, Heidelberg, 2015).

\bibitem{book} B. A. Malomed, \textit{Multidimensional Solitons} (American
Institute of Physics, Melville, NY, 2022).

\bibitem{solitons} N. J. Zabusky and M. D. Kruskal, Interaction of
\textquotedblleft solitons" in a collisional plasma and the recurrence of
initial states, Phys. Rev. Lett. \textbf{15}, 240-243 (1965).

\bibitem{Townes} Chiao R. Y., E. Garmire, and C. H. Townes, Self-trapping of
optical beams, Phys. Rev. Lett. \textbf{13}, 479-482 (1964).

\bibitem{virial} S. N. Vlasov, V. A. Petrishchev, and V. I. Talanov, Izv.
Vyssh. Uchebn. Zaved. Radiofiz. \textbf{14}, 1353-1363 (1971) [English
translation: Radiophys. Quantum Electron. \textbf{14}, 1062-1070 (1971)].

\bibitem{Anderson} M. Desaix, D. Anderson, and M. Lisak, Variational
approach to collapse of optical pulses, J. Opt. Soc. Am. B \textbf{8},
2082-2086 (1991).

\bibitem{TS-exper} B. Bakkali-Hassani B., C. Maury, Y.-Q. Zhou, E. Le Cerf,
R. Saint-Jalm, P. C. M. Castilho, S. Nascimbene, J. Dalibard, and J.
Beugnon, Realization of a Townes soliton in a two-component planar Bose gas,
Phys. Rev. Lett. \textbf{127}, 023603 (2021).

\bibitem{TS-exper 2} B. Bakkali-Hassani, C. Maury, S. Stringari, S.
Nascimbene, J. Dalibard, and J. Beugnon, The cross-over from Townes solitons
to droplets in a 2D Bose mixture, New J. Phys. \textbf{25}, 013007 (2023).

\bibitem{1D-TS} F. Kh. Abdullaev and M. Salerno, Gap-Townes solitons and
localized excitations in low-dimensional Bose-Einstein condensates in
optical lattices, Phys. Rev. A \textbf{72}, 033617 (2005).

\bibitem{1D-TS-2} G. L. Alfimov, V. V. Konotop, and P. Pacciani, Stationary
localized modes of the quintic nonlinear Schr\"{o}dinger equation with a
periodic potential \textbf{75}, 023624 (2007).

\bibitem{Cid-OptExp} A. S. Reyna and C. B. de Ara\'{u}jo, Spatial phase
modulation due to quintic and septic nonlinearities in metal colloids, Opt.
Exp. \textbf{22}, 22456-22469 (2014).

\bibitem{Cid} A. S. Reyna and C. B. de Ara\'{u}jo, High-order optical
nonlinearities in plasmonic nanocomposites -- a review, Adv. Opt. Phot.
\textbf{9}, 720-774 (2017).

\bibitem{Berge12} L. Berg\'{e}, J. Juul Rasmussen, and J. Wyller, Dynamics
of localized solutions to the Raman-extended derivative nonlinear Schr\"{o}%
dinger equation, J. Phys. A: Math. Gen. \textbf{29}, 3581-3595 (1996); J. S.
Hesthaveny, J. Juul Rasmussen, L. Berg\'{e}, and J. Wyller, Numerical
studies of localized wavefields governed by the Raman-extended derivative
nonlinear Schr\"{o}dinger equation, J. Phys. A: Math. Gen. \textbf{30},
8207-8224 (1997).

\bibitem{Hasegawa} Y. Kodama and A. Hasegawa, Nonlinear pulse propagation in
a monomode dielectric guide, IEEE J. Quant. Elect. \textbf{QE-23}, 510-524
(1987).

%\bibitem{TG1} G. E. Astrakharchik, J. Boronat, J. Casulleras, and S.
%Giorgini, Beyond the Tonks-Girardeau gas: Strongly correlated regime in
%quasi-one-dimensional Bose gases, Phys. Rev. Lett. \textbf{95}, 190407
%(2005).

%\bibitem{TG2} M. T. Batchelor, M. Bortz, X. W. Guan, and N. Oelkers,
%Evidence for the super Tonks-Girardeau gas, J. Stat. Phys. -- Theory and
%Experiment L10001 (2005).

\bibitem{VK} N. G. Vakhitov and A. A. Kolokolov, Stationary solutions of the
wave equation in a medium with nonlinearity saturation, Radiophys. Quantum
Electron. \textbf{16}, 783-789 (1973); https://doi.org/10.1007/BF01031343.

\bibitem{APL} S. R. Friberg, Y. Silberberg, M. K. Oliver, M. J. Andrejko, M.
A. Saifi, and P. W. Smith, Ultrafast all-optical switching in a dual-core
fiber nonlinear coupler, Appl. Phys. Lett. \textbf{51}, 1135-1137 (1987).

\bibitem{Wright} E. M. Wright, G. I. Stegeman, and S. Wabnitz, Solitary-wave
decay and symmetry-breaking instabilities in two-mode fibers. Phys. Rev. A
\textbf{40}, 4455-4466 (1989).

\bibitem{Snyder} A. W. Snyder, D. J. Mitchell, L. Poladian, D. R. Rowland,
and Y. Chen, Physics of nonlinear fiber couplers, J. Opt. Soc. Am. B \textbf{%
8}, 2101-2118 (1991).

\bibitem{Wabnitz} M. Romagnoli, S. Trillo, and S. Wabnitz, Soliton switching
in nonlinear couplers, Opt. Quantum Electron \textbf{24}, S1237--S1267
(1992).

\bibitem{review} B. A. Malomed, A variety of dynamical settings in dual-core
nonlinear fibers, In: Handbook of Optical Fibers, Vol. 1, pp. 421-474 (G.-D.
Peng, Editor: Springer, Singapore, 2019).

\bibitem{Smerzi} A. Smerzi, S. Fantoni, S. Giovanazzi, and S. R. Shenoy,
\textquotedblleft Quantum coherent atomic tunneling between two trapped
Bose-Einstein condensates", Phys. Rev. Lett. \textbf{79}, 4950-4953 (1997).

\bibitem{Markus} M. Albiez, R. Gati, J. F\"{o}lling, S. Hunsmann, M.
Cristiani, and M. K. Oberthaler, Direct observation of tunneling and
nonlinear self-trapping in a single bosonic Josephson junction, Phys. Rev.
Lett. \textbf{95}, 010402 (2005).

\bibitem{Bugar} V. H. Nguyen, L. X. T. Tai, I. Bugar, M. Longobucco, R.
Buzcynski, B. A. Malomed, and M. Trippenbach, Reversible ultrafast soliton
switching in dual-core highly nonlinear optical fibers, Opt. Lett. \textbf{45%
}, 5221-5224 (2020).

\bibitem{LiWang} L. Wang, B. A. Malomed, and Z. Yan, Attraction centers and $%
\mathcal{PT}$-symmetric delta-functional dipoles in critical and
supercritical self-focusing media, Phys. Rev. E \textbf{99}, 052206 (2019).

\bibitem{Azbel} B. A. Malomed and M. Ya. Azbel, Modulational instability of
a wave scattered by a nonlinear center, Phys. Rev. B \textbf{47},
10402-10406 (1993).

\bibitem{Kip} J. Hukriede, D. Runde, and D. Kip, Fabrication and application
of holographic Bragg gratings in lithium niobate channel waveguides, J.
Phys. D \textbf{36}, R1 (2003).

\bibitem{Feshbach1} R. Yamazaki, S. Taie, S. Sugawa, and Y. Takahashi,
Submicron spatial modulation of an interatomic interaction in a
Bose-Einstein condensate, Phys. Rev. Lett. \textbf{105}, 050405 (2010).

\bibitem{Feshbach2} L. W. Clark, L.-C. Ha, C.-Y. Xu, and C. Chin, Quantum
dynamics with spatiotemporal control of interactions in a stable
Bose-Einstein condensate, Phys. Rev. Lett. 115, 155301 (2015).

\bibitem{fused2} A. Takagi, K. Jinguji, and M. Kawachi, Wavelength
characteristics of ($2\times 2$) optical channel-type directional couplers
with symmetric or nonsymmetric coupling structures, J. Lightwave Tech.
\textbf{10}, 735-746 (1992).

\bibitem{fused1} R. Ismaeel, T. Lee, B. Oduro, and Y. Jung, and G.
Brambilla, All-fiber fused directional coupler for highly efficient spatial
mode conversion, Opt. Exp. \textbf{22}, 11610-11619 (2014).

\bibitem{Raymond} Y. Li, W. Pang, S. Fu, and B. A. Malomed, Two-component
solitons under a spatially modulated linear coupling: Inverted photonic
crystals and fused couplers, Phys. Rev. A \textbf{85}, 053821 (2012).

\bibitem{Alon} A. Harel and B. A. Malomed, Interactions of spatial solitons
with fused couplers, Phys. Rev. A \textbf{89}, 043809 (2014).

\bibitem{Chen} C.-L. Chen, \textit{Foundations for Guided-Wave Optics}
(Wiley-Interscience, Hoboken, NJ, 2006).

\bibitem{couplers} R. Halir, P. J. Bock, P. Cheben, A. Ortega-Monux, C.
Alonso-Ramos, J. H. Schmid, J. Lapointe, D. X. Xu, J. G. Wanguemert-Perez,
and I. Molina-Fernandez, Waveguide sub-wavelength structures: a review of
principles and applications, Laser \& Phot. Reviews \textbf{9}, 25-49 (2015).

\bibitem{Newton} J. Nocedal and S. J. Wright, \textit{Numerical Optimization}
(Springer, New York, 2006).

\bibitem{Yang} J. Yang, \textit{Nonlinear Waves in Integrable and
Nonintegrable Systems} (SIAM, Philadelphia, 2010).

\bibitem{bifurcations} G. Iooss and D. D. Joseph, \textit{Elementary
Stability and Bifurcation Theory} (Springer, Berlin, 1980).
\end{thebibliography}
\end{document}